\documentclass[a4paper,onecolumn,10pt,accepted=2026-08-06]{quantumarticle}
\pdfoutput=1
\usepackage[utf8]{inputenc}
\usepackage[english]{babel}
\usepackage[T1]{fontenc}
\usepackage{amsmath}
\usepackage{amssymb}
\usepackage{hyperref}
\usepackage{enumitem}
\usepackage[numbers,sort&compress]{natbib}
\usepackage{braket}
\usepackage{lineno}
\usepackage{cleveref}
\usepackage{tabularx}
\usepackage[dvipsnames]{xcolor}
\usepackage{tikz}
\usepackage{lipsum}
\usepackage{caption} % Ensure compatibility with caption package
\usepackage{subcaption}

\usetikzlibrary{positioning}
\usetikzlibrary{shapes.geometric, arrows}
\usetikzlibrary{matrix}

\usepackage{amsthm}
\usepackage{arydshln}

\DeclareMathOperator{\GL}{GL}
\DeclareMathOperator{\Sp}{Sp}

\newtheoremstyle{break}%
    {}{}%
    {}{}% 
    {\bfseries}{}%  % Note that final punctuation is omitted.
    {\newline}{}

\theoremstyle{break}
\newtheorem{proposition}{Proposition}[section]
\newtheorem{lemma}{Lemma}[section]

\theoremstyle{remark}

\usepackage{algpseudocode}
\usepackage{algorithm}

\usepackage{xspace}
\newcommand{\CNOT}{CNOT\xspace}

\newenvironment{added}{}{}

\begin{document}
\author{Mark Webster}
\affiliation{Department of Physics \& Astronomy, University College London, London, WC1E 6BT, United Kingdom}
\thanks{\ Corresponding author: \url{mark.acacia@gmail.com}}
\author{Stergios Koutsioumpas}
\affiliation{Department of Physics \& Astronomy, University College London, London, WC1E 6BT, United Kingdom}
\author{Dan E Browne}
\affiliation{Department of Physics \& Astronomy, University College London, London, WC1E 6BT, United Kingdom}

\title{Heuristic and Optimal Synthesis of CNOT and Clifford Circuits}
\maketitle
\begin{abstract}
Efficiently implementing Clifford circuits is crucial for quantum error correction and quantum algorithms.
Linear reversible circuits, equivalent to circuits composed of \CNOT gates, have important applications in classical computing.
In this work we present methods for \CNOT and general Clifford circuit synthesis which can be used to minimise either the entangling two-qubit gate count or the circuit depth.
We present three families of algorithms - optimal synthesis which works on small circuits, A* synthesis for intermediate-size circuits and greedy synthesis for large circuits.
We benchmark against existing methods, including rustiq, tket and qiskit and show that our approach results in circuits with lower two-qubit gate count. For encoding circuits, our methods outperform previous reinforcement learning results and find a lower two-qubit gate count circuit for the Golay code than previously known.
The algorithms have been implemented in a \href{https://github.com/m-webster/CliffordOpt}{GitHub repository} for use by the classical and quantum computing community.
\end{abstract}
%\tableofcontents

\section{Introduction}
The Clifford group, generated by the single-qubit phase ($S:=\sqrt{Z}$) and Hadamard ($H$) operators, plus the two-qubit controlled-not (\CNOT) operator, plays an important part in quantum algorithms and quantum error correction. 
Examples of Clifford circuits used in the field of quantum error correction include encoding circuits, syndrome extraction circuits and logical gates. 
Due to the importance of Clifford circuits, finding efficient implementations has been an active area of research. 
In the classical literature, there is a significant amount of work on synthesis of linear reversible circuits which correspond to Clifford circuits using only \CNOT gates \cite{peephole, Patel_Markov_Hayes_2008, schaeffer2014costminimizationapproachsynthesis, Brugiere1, Murphy_Kissinger_2023}. 
The efficiency of a circuit is usually measured in terms of either the total number of two-qubit entangling gates or the depth of the circuit (the number of time-steps required to implement the circuit). 
\begin{added}
    Optimising the gate-count and depth of Clifford circuits is challenging because each Clifford operator has many possible circuit implementations with very different gate-counts and depths.
\end{added}
 
In this work, we introduce heuristic (greedy and A*) and optimal algorithms for circuit synthesis. 
Our algorithms can be applied to either \CNOT circuits or general Clifford circuits.
We use a mapping from \CNOT circuits to invertible binary matrices in $\GL(n,2)$ and from general Clifford operators to symplectic binary matrices in $\Sp(2n,2)$.
We consider equivalence classes of circuit matrix representations up to input/output qubit permutations and this results in efficiency gains and lower two-qubit gate counts.
In several qubit architectures, qubit permutations can be implemented with low overhead \cite{2002_Kielpinski, Baart2015SinglespinC, 2024_rydberg_atoms, Photonic} and so we can choose to implement the member of an equivalence class with the lowest two-qubit gate-count or depth. 
In addition, we can often absorb qubit permutations into the state preparation part of a circuit, even where low-overhead SWAP gates are not available \cite{rodriguez2024experimentaldemonstrationlogicalmagic, fazio2025lowoverheadmagicstatecircuits}.

Our greedy algorithm chooses gates to apply by optimising a vector derived from the column sums of the binary matrix representative of the operator. 
Existing greedy synthesis algorithms for \CNOT circuits optimise a scalar heuristic but tend to become trapped in local minima \cite{schaeffer2014costminimizationapproachsynthesis} and our vector-based method is much less susceptible to this issue. 

Our optimal synthesis algorithm involves generating a database of matrices by applying all possible two-qubit gates in sequence.
Using a variation of the method in \cite{Bravyi_Latone_Maslov_2022}, we reduce the size of the database by considering equivalence classes of matrices up to row/column permutations, transpose, inverse and, for Clifford circuits, single-qubit Clifford operators applied to the left and right. 
To synthesise a circuit, we look up the equivalence class in the database and return the saved gate sequence. 
For optimal \CNOT synthesis, we determine the equivalence class of a binary invertible matrix up to row and column permutations by mapping it to a bi-coloured graph using the method in \cite{ZIVKOVIC2006310} and checking for graph isomorphisms using open source packages such as nauty \cite{MCKAY201494} or Bliss \cite{Bliss1,Bliss2}.
We show how to extend this method to find equivalence classes of symplectic matrices up to permutations and single-qubit Clifford gates acting on both the left and right.
This allows us to generate a significantly smaller database using less computational resources compared to the method in \cite{Bravyi_Latone_Maslov_2022}.

Our A* synthesis algorithm uses a scalar heuristic $h$ which estimates the number of gates required to reduce the matrix to a permutation matrix (in the case of \CNOT circuits) or a permutation combined with a series of single-qubit Clifford operators (for the general symplectic case). 
The algorithm does not commit to an action immediately, instead saving gate options to a priority queue.
By analysing the databases generated in optimal synthesis, we find robust estimators for $h$ based on the row and column sums of matrices. 
The resulting A* algorithm usually outperforms the greedy algorithm when applied to intermediate circuit sizes.

Benchmarking over a range of circuit sizes, we show that our algorithms have favourable entangling two-qubit gate counts compared to existing methods for both \CNOT and general Clifford circuits. 
Our algorithms have been implemented in a \href{https://github.com/m-webster/CliffordOpt}{Python package} available for use by the community. 

\section{\CNOT Circuit Synthesis}\label{sec:CNOT_synthesis}
In this section, we consider synthesis of circuits composed entirely of \CNOT gates. 
Circuits of this type have important applications in syndrome extraction and magic state distillation and are also of interest in classical computing applications.
The structure of this section is as follows. 
We first show how to represent any \CNOT circuit as an invertible binary matrix such that adding columns of the matrix corresponds to \CNOT gates and introduce \CNOT synthesis via Gaussian elimination.
We then present our three new algorithms for \CNOT synthesis.
We conclude the section by benchmarking our methods against existing methods, and giving a number of applications.

% \begin{figure}[H]
%      \centering
%         \includegraphics[width=0.75\textwidth]{images/CNOT_benchmark.pdf}
%         \caption{\CNOT synthesis algorithm benchmark. Using $400$ randomly generated binary invertible matrices for $3 \le n \le 64$, we compare our greedy,  optimal and A*  \CNOT synthesis algorithms of \Cref{sec:CNOT_greedy,sec:CNOT_optimal,sec:CNOT_astar} with the algorithms in \cite{Brugiere1} (BBVMA) and in \cite{Patel_Markov_Hayes_2008} (PMH). The \CNOT-count of the A* algorithm closely matches optimal gate counts where available and has the lowest \CNOT-counts on circuits of up to 32 qubits. 
%         % The greedy algorithm using the $H_\text{prod}$ heuristic of \Cref{eq:H_heuristics} gives the best results for $10 \le n \le 24$ qubits. 
%         For larger circuits, the greedy algorithm with the $\mathbf{h}$ vector heuristic has the lowest \CNOT-count. }
%         \label{fig:CNOT_benchmark}
% \end{figure}

\begin{figure}
     \centering
     \begin{subfigure}[t]{0.49\textwidth}
         \centering
    \includegraphics[width=\textwidth]{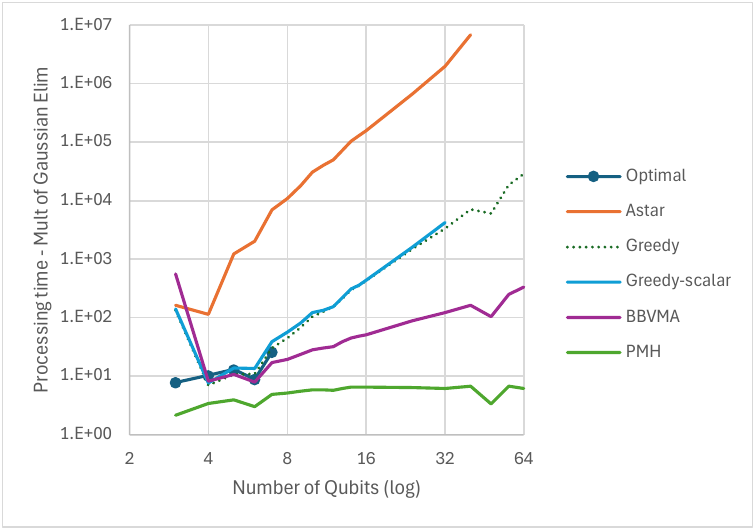}
    \caption{Processing Time }
     \end{subfigure}
     \hfill
     \begin{subfigure}[t]{0.49\textwidth}
         \centering
        \includegraphics[width=\textwidth]{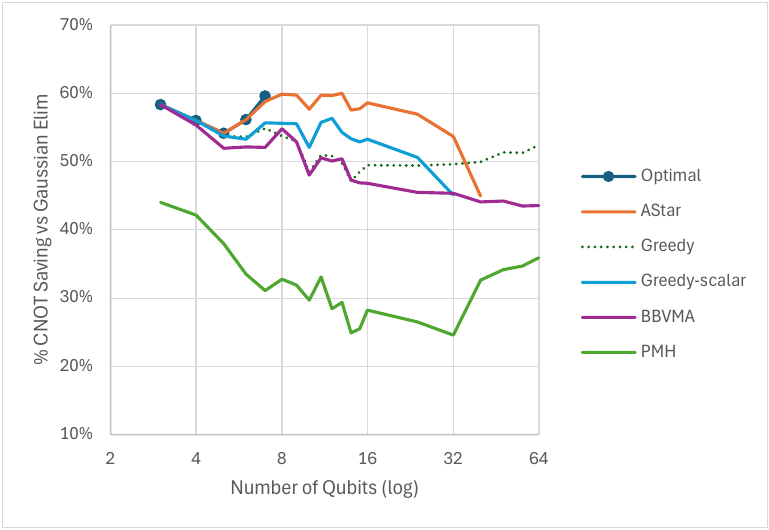}
        \caption{Two-Qubit Gate Count }
     \end{subfigure}
        \caption{\begin{added}\CNOT synthesis algorithm benchmark. Using $400$ randomly generated binary invertible matrices for $3 \le n \le 64$, we compare our greedy,  optimal and A*  \CNOT synthesis algorithms of \Cref{sec:CNOT_greedy,sec:CNOT_optimal,sec:CNOT_astar} with the algorithms in \cite{Brugiere1} (BBVMA) and in \cite{Patel_Markov_Hayes_2008} (PMH). The \CNOT-count of the A* algorithm closely matches optimal gate counts where available and has the lowest \CNOT-counts on circuits of up to 32 qubits. 
        For larger circuits, the greedy algorithm with the $\mathbf{h}$ vector heuristic has the lowest \CNOT-count.
        The processing time is expressed as a multiple of the time taken by Gaussian Elimination synthesis (lower is better). Two-qubit gate count is in terms of the percentage reduction in gate count from GE synthesis (higher is better). Both are plotted against the number of qubits in the circuit.\end{added}}
        \label{fig:CNOT_benchmark}
\end{figure}

\subsection{Representation of \CNOT Circuits as Invertible Binary Matrices}\label{sec:CNOT_representation}
In this section, we show how to represent a \CNOT circuit as an invertible binary matrix known as a parity matrix following the material in \cite{Murphy_Kissinger_2023}.
For any \CNOT circuit, we can find a parity matrix $A$ such that the action of the circuit on a computational basis vector $\ket{\mathbf{e}}$ is given by $\ket{\mathbf{e}A}$.
To find the parity matrix, we give $A$ the initial value $I_n$ and update $A$ by applying each \CNOT gate in sequence.
To apply a $\textrm{CNOT}_{ij}$ gate, we add column $i$ of $A$ to column $j$.
This is equivalent to multiplying on the right by a binary $n\times n$ matrix which is zero apart from the entry $(i,j)$ and the diagonal entries which are one.
% The SWAP gate $\textrm{SWAP}_{ij} = \textrm{CNOT}_{ij}\textrm{CNOT}_{ji}\textrm{CNOT}_{ij}$ is equivalent to swapping column $i$ of $A$ with column $j$ and is equivalent to multiplying on the right by a permutation matrix.

\subsection{CNOT Circuit Synthesis via Gaussian Elimination}\label{sec:gaussian}
Synthesis via Gaussian elimination employs the algorithm for determining the parity matrix introduced in \Cref{sec:CNOT_representation}, but operated in reverse. 
To perform circuit synthesis, we reduce the parity matrix for the desired \CNOT circuit to identity via Gaussian elimination by addition of columns (equivalent to a \CNOT gate). 
Applying the gates in reverse order yields a synthesis of the \CNOT circuit.

Where the qubit architecture allows SWAP gates to be implemented with low overhead, we can lower the \CNOT-count by reducing the parity matrix to a permutation matrix $P$ instead of the identity. 
% In this case, we are finding a \CNOT circuit for equivalence classes of parity matrices up to column permutations (i.e qubit qubit permutations).
Pseudocode for a  Gaussian elimination algorithm using this method can be found in \Cref{alg:gaussian}.

In \cite{Patel_Markov_Hayes_2008} the authors showed that any \CNOT circuit can be implemented by using $O(n^2/\log n)$ \CNOT gates.
Gaussian elimination gives a circuit using $O(n^2)$ \CNOT gates, and so the algorithm is not asymptotically optimal.
We will use Gaussian elimination as our benchmark when comparing algorithms for \CNOT synthesis.

\subsection{Greedy Synthesis of CNOT Circuits}\label{sec:CNOT_greedy}
In this subsection we present our greedy algorithm for \CNOT circuit synthesis which has improved \CNOT-count compared to Gaussian elimination.
The algorithm works by considering all possible \CNOT gates at each stage and applying the one which minimises a particular heuristic. 

Various greedy \CNOT synthesis algorithms have been proposed in previous works and we discuss these here.
Synthesis via Gaussian elimination guarantees at each step that at least one entry is eliminated by adding two columns. 
In practice, we can often eliminate more than one entry if there is a high degree of overlap between two columns.
The PMH algorithm of \cite{Patel_Markov_Hayes_2008} utilises this fact to increase the number of entries eliminated in each column operation. 
Each column is broken into a series of blocks and any blocks which are equal are added together first.
The authors show that the gate-count of the PMH algorithm matches an  asymptotically optimal figure of $O(n^2/\log n)$.

There are a number of works using the idea of choosing \CNOT gates which eliminate as many entries as possible.
In \cite{schaeffer2014costminimizationapproachsynthesis,Brugiere1} the authors consider algorithms which optimise the selection of \CNOT gates by minimising the number of ones in the parity matrix $A$:
\begin{align}
h_\text{sum}(A) &:= \left(\sum_{0 \le i,j<n}A_{ij}\right).
\end{align}
In \cite{Brugiere1}, the authors introduce a metric based on  the sum of the logarithms of the column sums of the parity matrix.
This method gives priority to eliminating entries in columns which are `almost done' and have weight close to 1:
\begin{align}
h_\text{prod}(A) &:= \sum_{0\le j <n}\log\left(\sum_{0 \le i < n}A_{ij}\right).
\end{align}
These works also consider the inverse and transpose of the parity matrix when calculating their heuristics because these can be synthesized using the same number of \CNOT gates and may have lower heuristics.

The main drawback of optimising for scalar quantities is that the algorithm tends to get trapped in local minima.
The authors of \cite{schaeffer2014costminimizationapproachsynthesis} address this by reverting to Gaussian elimination if this occurs. 

In \cite{fazio2025lowoverheadmagicstatecircuits}, MW in collaboration with Nicholas Fazio and Zhenyu Cai, introduced a vector-based heuristic for \CNOT circuit synthesis in the context of simplifying magic state distillation circuits.
In that method, we reduced the parity matrix to a permutation matrix which corresponds to a qubit permutation.
Qubit permutations can be moved to the beginning of the circuit and absorbed into state preparation. In this work, we refine the greedy algorithm in that paper and compare it to existing methods in the literature.

\begin{figure}
     \centering
     \begin{subfigure}[t]{0.32\textwidth}
         \centering
    \includegraphics[width=\textwidth]{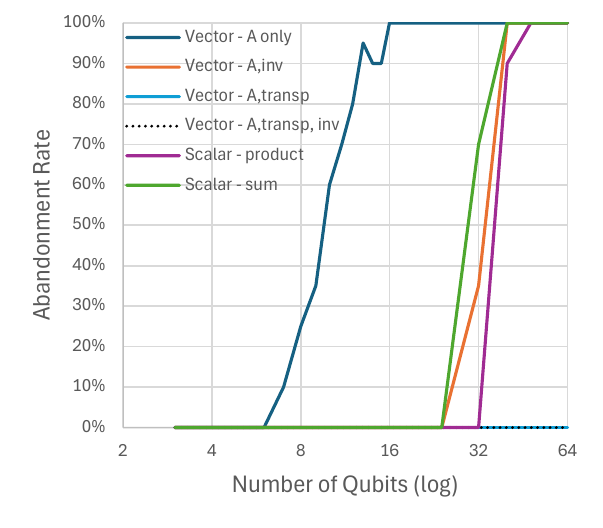}
    \caption{Abandonment Rate}\label{fig:CNOT_Heur_Aband}
     \end{subfigure}
     \begin{subfigure}[t]{0.32\textwidth}
         \centering
    \includegraphics[width=\textwidth]{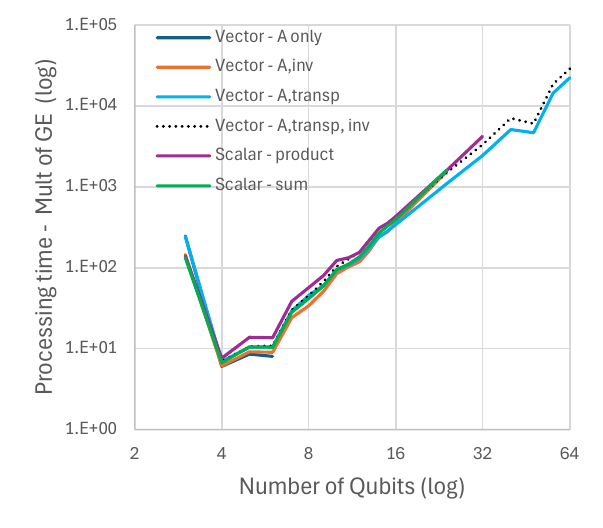}
    \caption{Processing Time }\label{fig:CNOT_Heur_Time}
     \end{subfigure}
     \hfill
     \begin{subfigure}[t]{0.32\textwidth}
         \centering
        \includegraphics[width=\textwidth]{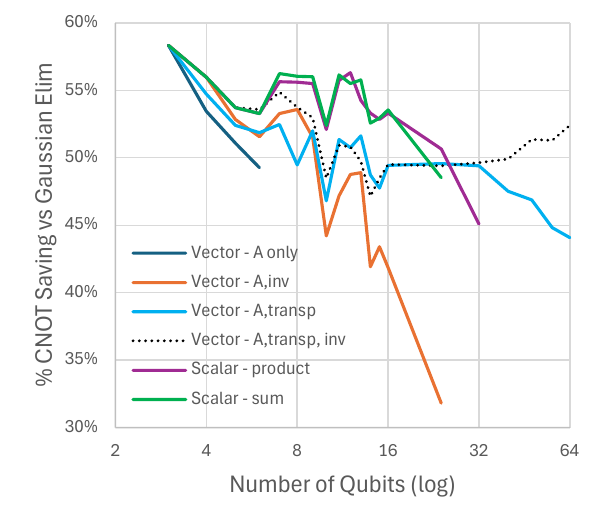}
        \caption{Two-Qubit Gate Count }\label{fig:CNOT_Heur_GateCount}
     \end{subfigure}
        \caption{\begin{added}Heuristics for greedy \CNOT synthesis. Using $400$ randomly generated binary invertible matrices for $3 \le n \le 64$, we compare the performance of greedy synthesis using various vector \Cref{eq:h_vector} and scalar heuristics of \Cref{eq:H_heuristics}. 
        When running the greedy algorithm, if the metric does not improve for 10 iterations, it is likely that it has become trapped in a local minimum and we abandon the run.
        In \Cref{fig:CNOT_Heur_Aband}, we plot the abandonment rate. The vector heuristic of \Cref{eq:h_vector} which uses the column sums of the matrix, its inverse and transpose has a zero abandonment rate for this data set. The scalar methods and the other vector methods all have an abandonment rate of 100\% for circuits with 64 qubits. 
        In \Cref{fig:CNOT_Heur_Time} we plot the processing time  as a multiple of the time taken by Gaussian Elimination synthesis (lower is better) and see that there is very little difference in processing time for the different heuristics. 
        In \Cref{fig:CNOT_Heur_GateCount}, we plot two-qubit gate count as a percentage reduction in gate count from Gaussian  synthesis (higher is better).
        We find that the scalar heuristics have lower \CNOT-count for $n \le 24$ qubits.
        The vector-based heuristic has a lower two-qubit gate count for  circuits with $32$ qubits or more.
        \end{added}}
        \label{fig:CNOT_greedy_heuristics}
\end{figure}

We show that our method is less likely to become trapped in local minima versus scalar heuristics. 
In our comparison, we use scalar heuristics based on those discussed in \cite{schaeffer2014costminimizationapproachsynthesis,Brugiere1} calculated from combinations of the transpose and inverse of the parity matrix, normalised so that they are zero for a permutation matrix:
\begin{align}
H_\text{sum}(A) &:= \left(h_\text{sum}(A) + h_\text{sum}(A^{-1}) \right)/2n-1,\nonumber\\
H_\text{prod}(A) &:= \left(h_\text{prod}(A) + h_\text{prod}(A^T) +h_\text{prod}(A^{-1})+h_\text{prod}(A^{-T})\right)/4n.\label{eq:H_heuristics}
\end{align}
Our vector heuristic is formed by sorting the column sums of the parity matrix, its inverse and its transpose.
We first calculate a vector of length $n$ which corresponds to sums of the columns of a matrix $A$.
We subtract $1$ from the column sums so that the vector is zero for permutation matrices.
The $j$th entry of the column sum vector is given by 
\begin{align}
    \text{colSums}(A)[j]&:= \sum_{0\le i < n}A_{ij} - 1
\end{align}
The vector heuristic is the result of sorting the length $4n$ vector formed by joining together the column sums of the parity matrix $A$, its transpose $A^T$, its inverse $A^{-1}$ and the inverse transpose $A^{-T}$:
\begin{align}
    \mathbf{h}(A) &:= \text{sorted}\left[\text{colSums}(A), \text{colSums}(A^T), \text{colSums}(A^{-1}), \text{colSums}(A^{-T})\right].\label{eq:h_vector}
\end{align}
\begin{added}
    The greedy algorithm using the new heuristic is described in full in \Cref{alg:greedy}.
    To calculate the column sums of the inverse and transpose, we update the symplectic matrix corresponding to the CNOT circuit during processing (see \Cref{sec:general_clifford_synth}). This includes both the parity matrix and inverse of the parity matrix and reduces the computational complexity of calculating the heuristic. 
\end{added}
The vector heuristic in \cite{fazio2025lowoverheadmagicstatecircuits} includes the column sums of the matrix and its transpose only.
\begin{added}
We tested the new vector heuristic using a dataset of 400 random binary invertible matrices with between 3 and 64 rows. For the new vector heuristic, we found that the greedy algorithm did not get trapped in local minima even once. On the other hand, alternative vector heuristics including column sums of the inverse or transpose or just the original matrix resulted in many such issues. In \Cref{fig:CNOT_Heur_Aband}, we summarise the results of the experiment.

% \begin{table}[]    \centering					
% \begin{tabular}{|	r|	r|	r|	r|	r|	}\hline
% \textbf{n}&	\textbf{A only}&	\textbf{A and Inverse}&	\textbf{A and Transpose}&	\textbf{A Transpose and Inverse}\\	
% \hline
% 4&	20&	20&	20&	20\\	
% 4&	20&	20&	20&	20\\	
% 5&	20&	20&	20&	20\\	
% 6&	20&	20&	20&	20\\	
% 7&	18&	20&	20&	20\\	
% 8&	16&	20&	20&	20\\	
% 9&	13&	20&	20&	20\\	
% 10&	9&	20&	20&	20\\	
% 11&	5&	20&	20&	20\\	
% 12&	3&	20&	20&	20\\	
% 13&	1&	20&	20&	20\\	
% 14&	2&	20&	20&	20\\	
% 15&	2&	20&	20&	20\\	
% 16&	0&	20&	0&	20\\	
% 24&	0&	20&	0&	20\\	
% 32&	0&	12&	0&	20\\	
% 40&	0&	0&	0&	20\\	
% 48&	0&	0&	0&	20\\	
% 56&	0&	0&	0&	20\\	
% 64&	0&	0&	0&	20\\	\hline
%     \end{tabular}					
%     \caption{Comparison of vector heuristics - completion rate. We selected 20 random invertible binary matrices of various sizes between 3 and 64 columns. We compared different vector heuristics which used: column sums of the matrix; column sums of the matrix and its inverse; column sums of the matrix and its transpose; and column sums of the matrix, its inverse transpose and transpose inverse. To test for cases where the greedy algorithm becomes trapped in local minima, we apply the algorithm and terminated early if the heuristic did not improve after 100 iterations. In this table, we list the number of cases where the algorithm completed without early termination.}		
%     \label{tab:vector_heuristic}					
% \end{table}					

\end{added}

We benchmarked our vector heuristic against the $H_\text{sum}$ and $H_\text{prod}$ scalar heuristics and tracked whether the algorithm became trapped in a local minimum.
For each algorithm run, we keep track of the minimum heuristic achieved so far. 
If the minimum heuristic has not improved after 10 iterations, we abandon the run on the basis that it has most likely become trapped in a local minimum.
Pseudocode for the greedy algorithm is set out in \Cref{alg:greedy} and includes the condition for abandoning the run.

We used a random dataset of 400 \CNOT circuits on up to 64 qubits for the benchmark and the results are plotted in \Cref{fig:CNOT_greedy_heuristics}.
The percentage saving of \CNOT-gates versus the Gaussian elimination method of \Cref{sec:gaussian} is used as our benchmark \begin{added}for two-qubit gate count. For processing time, Gaussian elimination will in general be much faster than the algorithms in our benchmark so we use the processing time of the algorithm divided by the processing time of Gaussian elimination for the same circuits.\end{added}.
We found that for all heuristics, the greedy algorithm is unlikely to become trapped in a local minimum for circuits on $n<32$ qubits.
In this range, the scalar heuristics $H_\text{sum}(A)$ and $H_\text{prod}(A)$ gave lower \CNOT-counts than the vector heuristic $\mathbf{h}(A)$. 
For larger circuits, the performance of the scalar heuristics deteriorated, with a $100\%$ abandonment rate for circuits on $n>32$ qubits.
For larger circuits, the vector-based heuristic was dominant and gave a 50\% reduction in \CNOT-count on average versus Gaussian elimination - the data is available on our \href{https://github.com/m-webster/CliffordOpt/blob/main/paper_results/GL_benchmark.xlsx}{GitHub repository}.

% \begin{figure}
%      \centering
%     \includegraphics[width=0.7\textwidth]{images/CNOT_greedy_heuristics.pdf}
%     \caption{Heuristics for greedy \CNOT synthesis. Using $400$ randomly generated binary invertible matrices for $3 \le n \le 64$, we compare the performance of greedy synthesis using the vector-based heuristic of \Cref{eq:h_vector} to the scalar heuristics of \Cref{eq:H_heuristics}. We find that the scalar heuristics have lower \CNOT-count for $n < 32$ but for $n \ge 32$ qubits, they tend to get trapped in local minima. The vector-based heuristic continues to perform well for large circuits. }
%     \label{fig:CNOT_greedy_heuristics}
% \end{figure}

\subsubsection{Greedy Depth Optimisation for \CNOT Circuits}
The methods discussed so far for greedy \CNOT synthesis focus on minimising the \CNOT count of circuits.
In this section, we show how to generalise the greedy algorithm to minimise the depth of the circuit.

Consider an iteration of the greedy algorithm where we consider all possible \CNOT gates.
In the depth-optimising version of the algorithm, we keep track of the heuristic $\mathbf{h}$ for the previous iteration.
For each gate option $G_i$, we calculate the depth $d_i$ of the resulting circuit.
If the heuristic $\mathbf{h}_i$ of the new gate is worse than $\mathbf{h}$, we add a large number to $d_i$ as a penalty.
We then select the \CNOT gate which minimises the vector $(d_i,\mathbf{h}_i)$, apply the gate $G_i$ and update $\mathbf{h}$.
Pseudocode for greedy depth optimisation is set out in \Cref{alg:greedy}.

\subsection{Optimal CNOT Circuit Synthesis}\label{sec:CNOT_optimal}
Whilst greedy \CNOT synthesis gives good results even for large circuits, it does not guarantee that the \CNOT-count will be optimal.
In this section we show how to find the minimal number of \CNOT gates required for synthesis of \CNOT circuits on a small number of qubits ($\le 7$ qubits) based on the method of \cite{Bravyi_Latone_Maslov_2022} for general Clifford circuits. 
During preparation of this paper, a work using a similar idea for \CNOT synthesis has been released using different methods and achieving different results \cite{christensen2025exactsizesminimalcnot}.
We generate a database of all equivalence classes of $\GL(n,2)$ up to permutations of the rows and columns plus transposes and inverses of matrices. 
Considering equivalence classes of $\GL(n,2)$ results in an exponential reduction in the  database size and processing time. 
The database is generated in order of the minimum \CNOT-count for each class. 
We also show how to generate the database in order of circuit depth, allowing for synthesis of minimum-depth circuit implementations of \CNOT circuits.
To synthesize a \CNOT circuit, we look up the equivalence class of the parity matrix in the database, then apply a combination of qubit permutations on the left and right of the circuit. 

\subsubsection{Generating Equivalence Classes of $\GL(n,2)$}\label{sec:OptDB}
To generate the matrix database, we start with the $n \times n$ identity matrix, then apply all possible $n(n-1)$ $\textrm{CNOT}_{ij}$ gates to find matrices requiring one \CNOT. 
To find matrices requiring $d+1$ \CNOT gates, we start with those requiring $d$ gates and again apply all possible \CNOT gates.

\begin{figure}
    \centering
    \includegraphics[width=0.6\linewidth]{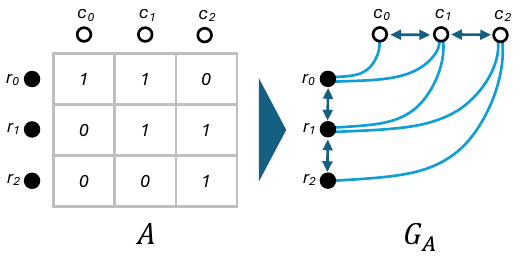}
    \caption{Mapping a binary invertible matrix $A$ to a bi-coloured graph $G_A$ using the method of \cite{ZIVKOVIC2006310}.
    Each row of the matrix is a vertex of type $r$ and each column  a  vertex of type $c$ in $G_A$. 
    Two vertices $r_i$ and $c_j$ are joined by an edge in $G_A$ if $A_{ij} = 1$.
    Graphs $G_A$ and $G_B$ are isomorphic if there is a permutation $P_r$ of rows and a permutation $P_c$ of the columns such that $(r_i,c_j)$ is an edge in $G_B$  $ \iff (P_r(r_i),P_c(c_j))$ is an edge in $G_A$. }
    \label{fig:GL_graph}
\end{figure}
We only add matrices to the database if they are not equivalent to an existing matrix up to permutations of the rows and columns and determining equivalence classes of matrices is the main technical challenge of this approach.
This question was addressed in \cite{ZIVKOVIC2006310} by representing a binary matrix $A$ as a bi-coloured graph $G_A$ as set out in \Cref{fig:GL_graph}. 
Two binary matrices are in the same equivalence class if and only if their graph representations are isomorphic where labels for vertices of the same type ($r$ or $c$) can be swapped. 
We use the nauty or Bliss package to test whether graphs are isomorphic \cite{MCKAY201494,Bliss1,Bliss2}. 
This involves generating a canonical labelling of the graph, which is a permutation of the row and column vertices which puts the graph into a canonical form. We calculate a certificate from the adjacency matrix by converting it into graph6 format string \cite{MCKAY201494}. The certificates for two graphs are equal if and only if the graphs are isomorphic. 
% The package also gives a canonical labelling of the graph and this can be used to generate the row and column permutations to map matrices in the equivalence class to each other. 

Once an invertible matrix has been synthesised, it is straightforward to obtain a synthesis for the transpose, the inverse and the transpose inverse using the same number of \CNOT gates. We only add the matrix to the database if the transpose, inverse and inverse transpose have not  been generated already.
Pseudocode for the database generation algorithm is set out in \Cref{alg:optimal_DB}.

We have generated all equivalence classes of invertible binary matrices up to $n=7$ and the results are summarised in \Cref{tab:GL2_classes}. 
We classify the equivalence classes of $\GL(n,2)$ by the minimal number of \CNOT gates and circuit depth in \Cref{tab:GL2_CNOT_count,tab:GL2_depth}. 
Even with the space savings resulting from considering equivalence classes of matrices, the size of the database grows exponentially with $n$.
\begin{added}
    It might well be possible to generate a database for $n=8$, though this is something we were not able to attempt due to capacity constraints on the university cluster. Due to the likely growth in database size, it is unlikely to be possible to generate a database for all circuits on $n > 8$ qubits. On the other hand, a partial database which includes circuits with less than a certain number of 2-qubit gates would require less storage and might well be useful.
\end{added}

\begin{table}[]
    \centering
    \begin{tabular}{| p{0.02\linewidth} | p{0.1\linewidth} | p{0.22\linewidth} |p{0.22\linewidth} |p{0.13\linewidth} |p{0.13\linewidth} |}
    \hline
n & $|\GL(n,2)|$ & Equiv classes up to col perms & Classes up to row/col perms, transp and inv &Max \CNOT-Count &Max Circuit Depth \\
\hline
1 & 1 & 1 & 1 & 0 & 0 \\
2 & 6 & 3 & 2 & 1  & 1\\
3 & 168 & 28 & 5 & 3  & 3\\
4 & 20160 & 840 & 27 & 6  & 3\\
5 & 1.00E+07 & 8.33E+04 & 284 & 8  & 5\\
6 & 2.02E+10 & 2.80E+07 & 11,761 & 12  & 5\\
7 & 1.64E+14 & 3.251E+10 & 1.72E+06 & 14 & 7 \\
\hline
    \end{tabular}
    \caption{Equivalence Classes of Binary Invertible Matrices. In this table, we compare the total number of invertible matrices $|\GL(n,2)|$ with the number of equivalence classes up to row and column permutations, plus transposes and inverses. The database for $n=7$ has $1.7$ million entries and uses around 34MB. 
    The maximum number of \CNOT gates required to generate any invertible matrix up to column and row permutations and the maximum depth of such a circuit are calculated as part of the optimal synthesis algorithm of \Cref{sec:CNOT_optimal} and are noted in the final two columns.}
    \label{tab:GL2_classes}
\end{table}

\subsubsection{Optimal Synthesis Algorithm}
The optimal synthesis algorithm takes as input an invertible binary matrix $A$. 
We then construct the bi-coloured graph $G_A$ and calculate the graph isomorphism class certificate as explained in \Cref{sec:OptDB}. 
We then search through the database to find a matrix $B$ with the same certificate. 
Using the canonical labellings of $A$ and $B$, we find row and column permutations $P_r, P_c$ such that $A = P_r B P_c$. 
We return the \CNOT gates used to generate $B$ conjugated by the permutations $P_r$ and $P_c$. 
Finally, the permutation $P_c$ is commuted through to the start of the circuit.

\subsubsection{Generalisation of Optimal Algorithm to Circuit Depth Optimisation}
The algorithm for generating the optimal circuit database can be modified to  yield minimum depth circuit implementations.
This is done by applying all possible depth-one \CNOT circuits at each step, rather than single \CNOT gates.
Generating all depth-one circuits on $n$ qubits is done by first generating all sets of one or more non-overlapping ordered pairs of qubits $(i,j)$, then applying all possible combinations of $\CNOT_{ij}$ and $\CNOT_{ji}$ to each pair.
We reduce algorithm complexity by only generating depth-one circuits which increase the circuit depth, and the algorithm for generating the sets of non-overlapping pairs of qubits is set out in \Cref{alg:depth-one}.

\subsection{A* Synthesis of CNOT circuits}\label{sec:CNOT_astar}
% Employing the  A* algorithm of \cite{hart_nilsson_astar} for \CNOT circuit synthesis yields an algorithm with lower two-qubit gate count than the greedy algorithm for circuits on up to $n=20$ qubits.
Optimal synthesis of \CNOT circuits via the method in \Cref{sec:CNOT_optimal} is only likely to be possible for up to $n=8$ qubits due to the size of the matrix database.
Here we describe an A* algorithm \cite{hart_nilsson_astar} which matches optimal results closely for $n\le 7$ and can be applied to circuits beyond this range. 
In the greedy algorithm of \Cref{sec:CNOT_greedy}, the gate which minimises the heuristic is applied at each step.
In the A* algorithm, gate options are stored as nodes in a priority queue rather than committing to a choice immediately.
At each stage, the node with the lowest value of $g(P)+h(P)$ is expanded, where $g(P)$ is the number of \CNOT gates applied  to reach the matrix $P$ and $h(P)$ is a heuristic which estimates the number of \CNOT gates required to reach a permutation matrix from $P$.
Providing the heuristic is admissible (i.e. does not over-estimate the true cost to reach the goal), the A* algorithm gives optimal results.

\subsubsection{Heuristic for A* Algorithm}
The main challenge of the A* method is to choose a suitable heuristic which estimates the number of \CNOT gates required to reach a permutation matrix.
Using the database generated for $n\le 7$ for the optimal \CNOT synthesis  method of \Cref{sec:CNOT_optimal}, we found a high correlation between the minimum \CNOT-count and the $H_\text{sum}$ heuristic of \Cref{eq:H_heuristics}, and a slightly higher correlation with $H_\text{prod}$ (\Cref{tab:correlation_GL}).
Accordingly, we used the heuristics $h = rH_\text{sum}(A)$ and $h=rH_\text{prod}(P)$ where $r$ is a parameter chosen to be close to the slope $m$ of the line of best fit as calculated in \Cref{tab:correlation_GL}.
We conducted a sensitivity analysis for both heuristics for a range of values for $r$. 
The data set used was a random set of 970 invertible $7\times 7$ matrices generated using the method of \Cref{sec:CNOT_optimal}.
% For each value of $d$, we selected up to 100 equivalence classes requiring $d$ \CNOT gates, or all equivalence classes if there were fewer than 100.
% Matrix representatives of the equivalence classes were `scrambled' by applying random row and column permutations.
For the $\GL(7,2)$ data set, we found that using the $H_\text{prod}$ heuristic gave more robust results and that varying $r$ may be necessary to achieve best results (see \Cref{fig:CNOT_astar_sensitivity}).
\begin{table}[]
    \centering
    \begin{tabular}{|l|r|r|r|r|r|r|}
    \hline
$H_\text{sum}(A)$ & n=2 & 3 & 4 & 5 & 6 & 7 \\
\hline
 R & 1.00 & 0.98 & 0.93 & 0.90 & 0.85 & 0.81 \\
 m & 2.00 & 2.22 & 2.44 & 2.36 & 2.25 & 2.20 \\
 b & 0.00 & 0.20 & 0.71 & 1.68 & 3.06 & 4.62 \\
\hline
\hline
$H_\text{prod}(A)$& n=2 & 3 & 4 & 5 & 6 & 7 \\
 \hline
 R & 1.00 & 0.98 & 0.93 & 0.91 & 0.87 & 0.84 \\
 m & 2.89 & 3.60 & 4.50 & 5.04 & 5.64 & 6.50 \\
 b & 0.00 & 0.20 & 0.52 & 1.13 & 1.89 & 2.53 \\
 \hline
    \end{tabular}
    \caption{Correlation between \CNOT-count and  heuristics for invertible $n\times n$ matrix $A$. We first calculate the correlation coefficient $R$, slope $m$ and intercept $b$ of the line of best fit for $H_\text{sum}(A)$ which uses the sum of the entries of $A$ and its inverse. We then do the same for  $H_\text{prod}(A)$  which uses the log of the row and column sums of $A$ and its inverse.}
    \label{tab:correlation_GL}
\end{table}

\subsubsection{Computational Complexity of A* Algorithm}
The main disadvantage of the A* algorithm is that it has exponential worst-case complexity, and the priority queue can become very large.
In our A* algorithm we manage the growth of the priority queue as follows.
Firstly, we only consider \CNOT operations between columns which have some overlap (and hence are likely to reduce the overall weight of the matrix).
Secondly, we maintain a maximum queue size at each step using the \href{https://pypi.org/project/treap/}{Treap python implementation} of Daniel Stromberg \cite{stromberg_treap}.
If the queue exceeds the size limit, we remove the  maximal entries.
We found that shortening the queue length can reduce the processing time for A*.
% - for instance we use a queue length of 100 for \Cref{fig:CNOT_benchmark} up to 15 qubits and a queue length of 10 thereafter.

\subsubsection{Generalisation of A* Algorithm to Circuit Depth Optimisation}
We now show how to generalise the A* algorithm to search for optimal depth circuit implementations.
This is done by using a modified $g$ metric.
Instead of counting the number of steps the algorithm takes to reach the matrix $P$, we set $g$ to be the depth of the \CNOT circuit required to reach $P$.
We obtained good results without altering the $h$ metric, though this could be the subject of future work.
Pseudocode for the A* algorithm which incorporates an option for depth optimisation is set out in \Cref{alg:astar_synthesis}.

\subsection{\CNOT Synthesis  Algorithm Benchmark}
In this section we compare our greedy, optimal and A*  \CNOT synthesis algorithms of \Cref{sec:CNOT_greedy,sec:CNOT_optimal,sec:CNOT_astar} to the BBVMA and  PMH algorithms \cite{Brugiere1,Patel_Markov_Hayes_2008}. 
The dataset used was the same set of 400 randomly generated binary invertible matrices in the range $3 \le n \le 64$ as in \Cref{fig:CNOT_greedy_heuristics}. 

We implemented both the BBVMA and PMH algorithms in python and these are available in our \href{https://github.com/m-webster/CliffordOpt}{GitHub repository} \cite{webster_m-cliffordopt}. 
The results of the PMH algorithm matched those in \cite{Patel_Markov_Hayes_2008} quite closely. 
The BBVMA algorithm first reduces the parity matrix to a triangular form which is then reduced to the identity. 
The authors focus mainly on the second step and demonstrate a fast algorithm which has a low \CNOT-count. 
We implemented the pseudocode of Algorithm 1 on page 7 of \cite{Brugiere1} to reduce a triangular matrix to identity.
We implemented a `minimising cost' algorithm for reducing a matrix to triangular form as described on page 13 of \cite{Brugiere1}. 
Our method chooses an $(i,j)$ pair at each step which results in the lowest $\CNOT$-count or the lowest matrix sum if the $\CNOT$-count is equal. 

% We compare our greedy, optimal and A*  \CNOT synthesis algorithms of \Cref{sec:CNOT_greedy,sec:CNOT_optimal,sec:CNOT_astar} to the BBVMA and  PMH algorithms \cite{Brugiere1,Patel_Markov_Hayes_2008}. 

The results of our \CNOT synthesis benchmark are in \Cref{fig:CNOT_benchmark}.
We found that the \CNOT-count of the A* algorithm matched the optimal number up to circuits on up to $n=7$ qubits where optimal data was available. 
To reduce the run-time of the A* algorithm, we used a relatively maximum low queue size of 100.
% for $n< 16$ and 10 thereafter.
We used the scaling heuristic $r=3$. The A* algorithm gave good results for circuits on up to $n=32$. 

For circuits on $n \ge 32$ the greedy algorithm with the $\mathbf{h}$ vector heuristic has the lowest \CNOT-count by an increasing margin.
A spreadsheet collating these results and giving the circuit implementations is available on our \href{https://github.com/m-webster/CliffordOpt/blob/main/paper_results/GL_benchmark.xlsx}{GitHub repository}.

\section{General Clifford Synthesis}\label{sec:general_clifford_synth}
In this section, we consider synthesis of general Clifford circuits.
Circuits of this type are used in quantum algorithms, in encoding circuits for quantum error correction and in logical gates for quantum codes.
In \cite{Aaronson_Gottesman_2004} Aaronson and Gottesman show that any Clifford operator has a canonical circuit implementation in 11 layers of the form:
$H-\textit{CNOT}-S-\textit{CNOT}-S-\textit{CNOT}-H-S-\textit{CNOT}-S-\textit{CNOT}$.
They show that AG synthesis is asymptotically optimal with $O(n^2/\log n)$ two-qubit gates.
In many cases far more efficient implementations exist, and finding these is the focus of this section.
We use AG synthesis as our benchmark for comparing algorithms.

Our approach is to use a circuit structure with only three layers - qubit permutations, single-qubit Cliffords, then 2-qubit symplectic transvections, which are equivalent to \CNOT gates up to single qubit Cliffords.
This structure allows us to simplify algorithms by considering equivalence classes of operators up to permutations and single-qubit Cliffords acting on either the right or the left hand side of the operator.
Transvections are quite natural gates for many quantum architectures, and give us much more freedom to optimise circuits compared to using just \CNOT gates \cite{Volanto, Pllaha_Volanto_Tirkkonen_2021}.
% We show how to modify the classical \CNOT synthesis methods introduced in \Cref{sec:CNOT_synthesis} for use in Clifford synthesis. This unifies the approaches and even allows us to re-use code in many cases.

The structure of this section is as follows. 
We first show how to represent Clifford operators as binary symplectic matrices and introduce symplectic transvections.
% We then introduce a modified version of the Volanto algorithm of \cite{Volanto} which synthesises Clifford operators using a series of two-qubit transvections. 
We then show how to generalise the greedy, optimal and A* algorithms of \Cref{sec:CNOT_greedy,sec:CNOT_optimal,sec:CNOT_astar} to general  Clifford circuits.
This unifies the approaches and even allows us to re-use code in many cases.
Finally, we benchmark our algorithms against existing methods for Clifford circuit synthesis.

% \begin{figure}[H]
%     \centering
%     \includegraphics[width=0.7\linewidth]{images/sym_benchmark.pdf}
%     \caption{Clifford synthesis algorithm benchmark. We used a dataset of 400 randomly generated symplectic matrices using the method of \cite{9605330}. The y-axis is the percentage reduction in two qubit gates compared to the Qiskit implementation of AG synthesis \cite{Aaronson_Gottesman_2004}.}
%     \label{fig:sym_benchmark}
% \end{figure}

\begin{figure}
     \centering
     \begin{subfigure}[t]{0.49\textwidth}
         \centering
    \includegraphics[width=\textwidth]{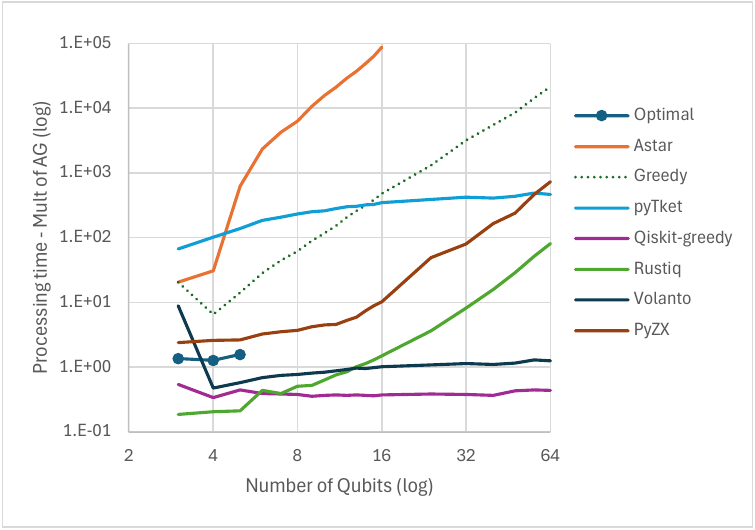}
    \caption{Processing Time }
     \end{subfigure}
     \hfill
     \begin{subfigure}[t]{0.49\textwidth}
         \centering
        \includegraphics[width=\textwidth]{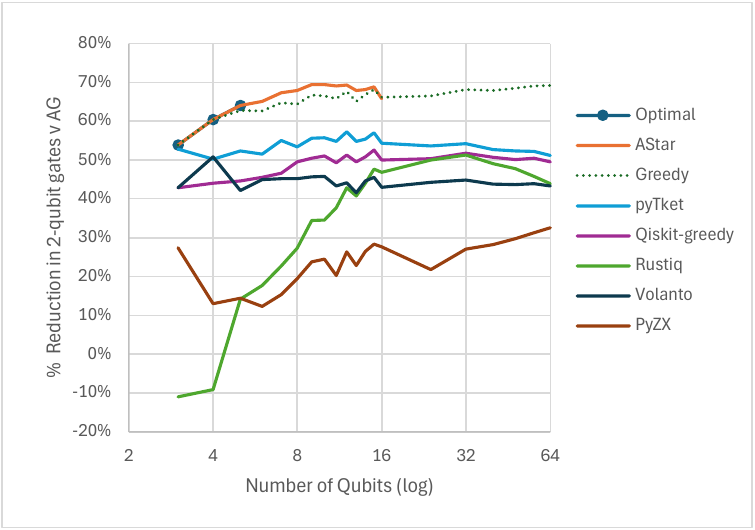}
        \caption{Two-Qubit Gate Count }
     \end{subfigure}
        \caption{\begin{added}Clifford synthesis algorithm benchmark. We used a dataset of 400 randomly generated symplectic matrices using the method of \cite{9605330}. 
        The processing time is expressed as a multiple of the time taken by Qiskit's Aaronson and Gottesman (AG) synthesis (lower is better). Two-qubit gate count is in terms of the percentage reduction in gate count from AG synthesis (higher is better). Both are plotted against the number of qubits in the circuit. We see that the A* algorithm is close to optimal for circuits on 5 or less qubits and has the lowest two-qubit gate count for circuits on up to 16 qubits. Thereafter, the greedy algorithm has a significantly lower gate count compared to the next best method, pyTket.\end{added}}
        \label{fig:sym_benchmark}
\end{figure}

\subsection{Binary Symplectic Representation of Clifford Operators}

In this subsection, we set out the binary symplectic representation of Clifford operators which will be used in our algorithms.
We represent Pauli operators as binary strings $(\mathbf{x}|\mathbf{z})$ where $\mathbf{x}$ and $\mathbf{z}$ are length $n$ binary vectors representing the $X$ and $Z$ components of the Pauli operator $i^{\mathbf{x}\cdot \mathbf{z}}X(\mathbf{x})Z(\mathbf{z})$ where $\mathbf{x}\cdot \mathbf{z}$ is the dot product of vectors and  $X(\mathbf{x}):=\prod_{0\le i <n}X_i^{\mathbf{x}[i]}$. 

We represent Clifford operators as $2n \times 2n$ binary symplectic matrices $A$ with four $n\times n$ sub-matrices:
\begin{align}
    A:=\left[\begin{array}{c|c}A_{XX}&A_{XZ}\\\hline A_{ZX}&A_{ZZ}\end{array}\right].
\end{align}\label{eq:Sp_4x4}
Clifford operators  map Paulis to Paulis under conjugation and are defined up to products of Paulis and global phases by specifying the mapping of the Pauli group generators $X_i$ and $Z_i$ for $0 \le i < n$.
Row $i$ of the matrix $[A_{XX}|A_{XZ}]$ indicates where the Clifford operator maps $X_i$ under conjugation. Similarly, row $i$ of $[A_{ZX}|A_{ZZ}]$ (which is row $i+n$ of $A$) gives the mapping of $Z_i$.
The matrices $A_{XX},A_{ZX}$ are the $X$-components of the mapped Paulis and  $A_{XZ},A_{ZZ}$ are the $Z$-components.
The Pauli operator represented by row $i$ commutes with all rows of $A$, apart from row $i+n$ with which it anti-commutes (see \cite{Aaronson_Gottesman_2004}).
\begin{added}
For instance, the single qubit Clifford Hadamard operator $H$ maps $X$ to $Z = HXH^{-1}$ and vice-versa under conjugation. 
Using the notation $\text{Sp}(H)$ for the symplectic matrix representation of the Hadamard operator, we have:
\begin{align}
    \Sp(H) = \left[\begin{array}{c|c}0&1\\\hline 1&0\end{array}\right].
\end{align}
Similarly, the $S$ operator maps $X$ to $Y = SXS^{-1} = iXZ$ under conjugation and leaves $Z$ unchanged. Hence, its symplectic representation is:
\begin{align}
    \Sp(S) = \left[\begin{array}{c|c}1&1\\\hline 0&1\end{array}\right].
\end{align}
The $S^{-1} = S^3 = SZ$ operator  has the same symplectic representation because it maps $X$ to $-Y = -iXZ$ under conjugation.
Hence we say that the symplectic representation is up to phases and Pauli products.
\end{added}
\subsection{Symplectic Transvections}\label{sec:transvections}
% A transvection is a geometrical shearing transformation. 
For synthesis of general Clifford circuits, we use transvections for our two-qubit gates.
Transvections play an important role in symplectic geometry, where it is known that they generate the symplectic group $\Sp(2n,2)$ \cite{o1978symplectic}, and correspond to geometrical shearing transformations.
When the Clifford group is represented using symplectic matrices, certain Clifford group gates can be identified as transvections \cite{Koenig_Smolin_2014,Rengaswamy_Calderbank_Pfister_Kadhe_2018}.
A transvection $ T_P$ on $n$ qubits is defined  in terms of a $n$-qubit Pauli operator $P$ as follows:
\begin{align}
    T_P := \exp\left(\frac{i\pi}{4}(I - P)\right).\label{eq:transvection_def}
\end{align}
It can easily be shown that $T_P^2 = P$ so we can think of $T_P = \sqrt{P}$. 
If $P$ has binary representation $\mathbf{v}:= (\mathbf{x}|\mathbf{z})$ then $T_P$ has the symplectic representation $I + \Omega\mathbf{v}^T\mathbf{v}$ \begin{added} where $\Omega := \begin{bmatrix}0&I\\I&0\end{bmatrix}$ is the binary symplectic form\end{added} (see \Cref{sec:transvection_proof}).
Two-qubit transvections have support on only two qubits, and are of the form:
\begin{align}
    \sqrt{P_iP_j} := T_{P_iP_j} = \exp\left(\frac{i\pi}{4}(I - P_iP_j)\right),\label{eq:2-qubit-transvections}
\end{align}
where $P_i, P_j$ are non-trivial single-qubit Paulis acting on qubits $i, j$ respectively.
As there are three non-trivial Paulis on a single qubit, there are $3 \times 3=9$ distinct two-qubit transvections. 

The Clifford group is generated by two-qubit transvections and single-qubit Cliffords, and we can write the \CNOT and CZ operators in terms of transvections as follows:
\begin{align}
    \textrm{CZ}_{ij} &= S_i^3 S_j^3 \sqrt{Z_iZ_j};\\
    \textrm{CNOT}_{ij} &= S_i^3\sqrt{X_j}^3\sqrt{Z_iX_j}.
\end{align}
\begin{added}
    These identities allow us to map the nine distinct transvections to CNOT operators conjugated by single-qubit Cliffords as used in \cite{Bravyi_Latone_Maslov_2022}. 
\end{added}
Two-qubit transvections are natural gates in a number of qubit architectures, including trapped ions\cite{trapped_ions}, NMR \cite{NMR}, spin qubits and quantum dot systems \cite{quantum_dots}  which further justifies this choice of gate set.
Up to a global phase, they are the same as the Pauli product rotations of \cite{Litinski_2019} with angle $\varphi = \pi/4$. 

\begin{added}
\subsection{Update Rules for Transvection Gate Set on Symplectic Matrices}
In this section we describe the action of our  gate set on symplectic matrices and discuss how to correct for possible sign errors. In \cite{Aaronson_Gottesman_2004} the authors explain how single-qubit Cliffords, CZ and CNOT operators update the tableau for a stabiliser code. As in the case of a symplectic matrix, each row of a stabiliser tableau is a Pauli string and it suffices to specify rules for how each operator updates Pauli strings under conjugation as well as any phase which is applied. 

In this work, our compilation gate set for Clifford operators comprises two-qubit transvections, single-qubit Cliffords generated by $H$ and $S$ and qubit permutations generated by qubit transpositions. 
Let $P$ and $Q$ be Pauli operators.
The action of the transvection $T_P$ acting via conjugation on the Pauli operator $Q$ is as follows (see \Cref{sec:transvection_proof}):
\begin{align}
    T_PQT_P^{-1} = \begin{cases}
        iPQ\text{: if }P, Q \text{ anti-commute; and} \\
        Q\text{: otherwise}.
    \end{cases}  
\end{align}
The Paulis $P$ and $Q$ commute if and only if the group commutator is $+1$. The group commutator can be calculated from the vector representations $P := X(\mathbf{p}_X)Z(\mathbf{p}_Z)$  and $Q := X(\mathbf{q}_X)Z(\mathbf{q}_Z)$ as follows:
\begin{align}
    PQP^{-1}Q^{-1} = -1^{\mathbf{p}_X \cdot \mathbf{q}_Z + \mathbf{p}_Z \cdot \mathbf{q}_X}
\end{align}
If the Paulis anti-commute, we update $Q$ by multiplying on the right by $P$ and applying a phase of $i$. The phase can be tracked using a phase vector which is an integer modulo $4$ representing the exponent of $i$ for each row of the symplectic matrix.

Update rules for the $S$ and $H$ operators are given in \cite{Aaronson_Gottesman_2004} and we describe these for a single-qubit Pauli $Q := X(q_X)Z(q_Z)$ where $q_X$ and $q_Z$ have values in $\{0,1\}$.
The action of $S$ acting via conjugation on $Q$ is given by:
\begin{align}
    SQS^{-1} = i^{q_X}X(q_X)Z(q_Z \oplus q_X).
\end{align}
The action of $H$ acting via conjugation on $Q$ is given by:
\begin{align}
    HQH = (-1)^{q_Xq_Z}X(q_Z)Z(q_X).
\end{align}
A qubit transposition (SWAP) on qubits $i$ and $j$ does not apply a phase, but swaps entry with index $i$ with the entry with index $j$ in both the X and Z-components of $Q$.

After finding a sequence of gates which synthesise a Clifford operator it is often the case that the phases generated by the above update rules is different to the default phase of a Pauli string $Q = i^{\mathbf{q}_X\cdot \mathbf{q}_Z}X(\mathbf{q}_X)Z(\mathbf{q}_Z)$ in one of the rows of the symplectic matrix. 
In this case, the phase can only differ by a factor of $\pm 1$ and we can correct the sign by prepending either a Pauli $Z$ (top half of the symplectic matrix) or Pauli $X$ (bottom half) to the circuit compilation.
For more details on Pauli corrections, see \cite{autQEC}.
\end{added}

\subsection{Greedy Synthesis of Clifford Circuits}\label{sec:clifford_greedy}
We now show how to modify the greedy \CNOT synthesis algorithm of \Cref{sec:CNOT_greedy} so that it can be applied to general Clifford circuits. 
% This results in an improved two-qubit gate-count than the modified Volanto algorithm at the expense of longer processing times. 
The greedy \CNOT circuit algorithm reduces the parity matrix to a permutation matrix.
The greedy Clifford algorithm reduces the symplectic matrix for the circuit to a matrix representing a qubit permutation followed by a layer of single-qubit Clifford gates. 
The termination condition is defined in terms of certain $2 \times 2$ sub-matrices of symplectic matrices:
\begin{align*}
    F_{ij} := 
    % \left[\begin{array}{c|c}A_{XX}_{i,j}&A_{XZ}_{i,j}\\\hline A_{ZX}_{i,j}&A_{ZZ}_{i,j}\end{array}\right] = 
    \left[\begin{array}{c|c}A_{i,j}&A_{i,j+n}\\\hline A_{i+n,j}&A_{i+n,j+n}\end{array}\right].\label{eq:Fij}
\end{align*}
The $F_{ij}$ are either rank 1, rank 2 (invertible) or all zero (rank 0). 
We keep track of the rank 1 and rank 2 $F_{ij}$ matrices by calculating the matrices $\textit{R1}(A)$ and $\textit{R2}(A)$ respectively as follows:
\begin{align}
    \textit{R1}(A)_{ij} &:= \begin{cases}
        1: \text{ if }\text{Rk}(F_{ij}) = 1;\\
        0: \text{ otherwise.}
    \end{cases}\\
    \textit{R2}(A)_{ij} &:= \begin{cases}
        1: \text{ if }\text{Rk}(F_{ij}) = 2;\\
        0: \text{ otherwise.}
    \end{cases}\label{eq:R12}
\end{align}
The greedy algorithm applies transvections to reduce the symplectic matrix to a form $P$ such that $\textit{R2}(P)$ is a permutation matrix and $\textit{R1}(P)$ is an all-zero matrix.
The permutation matrix $\textit{R2}(P)$ corresponds to a qubit permutation, and the rank 2 $F_{ij}$ in $P$ correspond to single-qubit Clifford gates. 
As $\textit{R2}(P)$ is a permutation matrix, there are  $n$ rank 2 $F_{ij}$, with exactly one in each row and column.
There are six invertible $2\times 2$ matrices, and these correspond to the symplectic representations of the six single-qubit Cliffords up to Pauli and phase factors.
This reduction method was inspired by the synthesis algorithm of Volanto \cite{Volanto}, which is analogous to Gaussian elimination for Clifford circuits. 
We describe a modified Volanto algorithm in \Cref{alg:volanto_synthesis} which reduces symplectic matrices to the form $P$ described above.

The greedy Clifford synthesis algorithm considers all possible pairs of qubits $(i,j)$ and all nine possible two-qubit transvections at each step.
It chooses the option which minimises the vector $\mathbf{h}(A)$ which is defined in a similar way to \Cref{eq:h_vector} using the column sums of $\textit{R2}$ and $\textit{R1}$ matrices of \Cref{eq:R12}:
\begin{align}
    \mathbf{h}(A) &:= \text{sorted}\left(\text{colSums}(\textit{R2}(A)+\textit{R1}(A)/n), \text{colSums}(\textit{R2}(A^T)+\textit{R1}(A^T)/n)\right).
\end{align}
We found that giving a lower weight to the column sums of the $\textit{R1}$ matrix gave good results, and in our metric we apply a factor of $n$ as the maxiumum column sum of an $\textit{R1}$ matrix is $n-1$.
Because the inverse of a symplectic matrix is given by $A^{-1} = \Omega A^T \Omega$, $\textit{R1}(A^{-1}) = \textit{R1}(A^{T})$  we do not need to consider the inverse when calculating the heuristic.

\subsection{Optimal Clifford Synthesis}\label{sec:clifford_optimal}
We now generalise the method of \Cref{sec:CNOT_optimal} by applying two-qubit symplectic transvections rather than \CNOT gates (see \Cref{eq:2-qubit-transvections}). 
Our method also can be used to perform depth-optimal searches by applying all possible depth one transvection circuits using the qubit partitioning algorithm of \Cref{alg:depth-one}.

\subsubsection{Generating Equivalence Classes of $\Sp(2n,2)$}
We consider equivalence classes of symplectic matrices up to multiplication on the left and right by symplectic matrices  corresponding to qubit permutations and a set of single qubit Clifford gates (see \Cref{sec:clifford_greedy}).
We have based our method on the one in \cite{Bravyi_Latone_Maslov_2022} where the authors show how to calculate a database of optimal circuits for Cliffords on up to 6 qubits.
Compared to \cite{Bravyi_Latone_Maslov_2022}, we generate a significantly smaller number of equivalence classes because we consider permutations of both the input and output qubits, and we also include transpose and inverses of matrices in the same class. 

\begin{figure}
    \centering
    \includegraphics[width=0.9\linewidth]{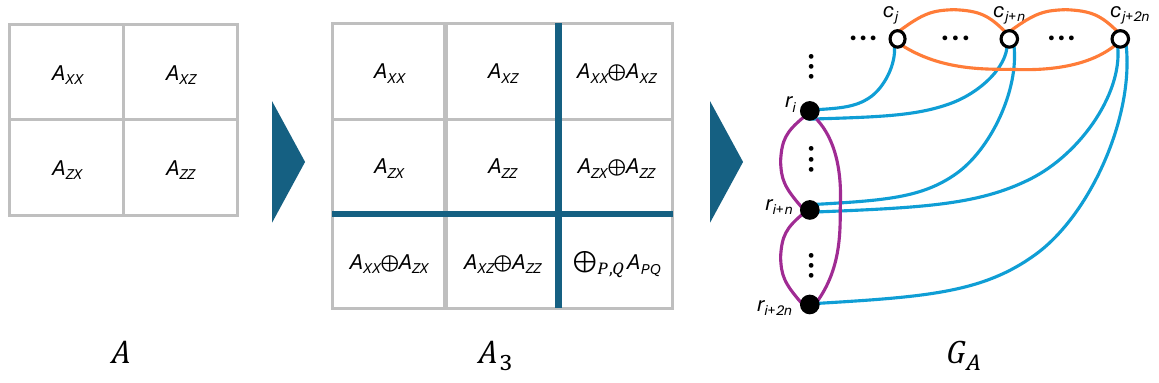}
    \caption{Mapping of symplectic matrices to bi-coloured graphs.
We first map the $2n \times 2n$ symplectic matrix $A$ of \Cref{eq:Sp_4x4} to a $3n \times 3n$ matrix $A_3$. 
This is done by adding a third block of rows and columns by summing across the blocks of $A$.
There is a vertex of type $r$ in $G_A$ for each row of $A_3$, and a vertex of type $c$ for each column.
There is an edge $(r_i,c_j)$ in $G_{A}$ if and only if the $i,j$ entry of $A_3$ is $1$.
We also add the edges $(r_i,r_{i+n}),(r_i,r_{i+2n})$ and $(r_{i+n},r_{i+2n})$ for $0 \le i < n$, then do the same for columns.
A permutation of the rows which preserves the connectivity of $G_A$ can be identified as a combination of SWAP and single-qubit Clifford gates applied on the left (see \cite{Bravyi_Latone_Maslov_2022}). 
Similarly, permutations of columns can be identified as permutations and single-qubit Cliffords applied on the right (see \cite{autQEC}).
}
    \label{fig:sym_graph}
\end{figure}

To determine if symplectic matrices are in the same equivalence class, we map to a graph isomorphism problem using a generalisation of the method in \Cref{sec:CNOT_optimal} and this is set out in \Cref{fig:sym_graph}. 
\begin{added}
    A proof that the Clifford operators $A$ and $B$ are equivalent up to single-qubit Cliffords and permutations acting on the left and right if and only if the graphs  $G_A$ and $G_B$ are isomorphic is set out in \Cref{sec:graph_iso_proof}.
\end{added}
We use the nauty \cite{MCKAY201494} or Bliss \cite{Bliss1,Bliss2} package to determine if two graphs $G_A$ and $G_B$ are isomorphic and to generate canonical labellings.

In \Cref{tab:Sp2_classes}, we summarise the number of equivalence classes of symplectic matrices on up to $n=5$ qubits and compare the number of equivalence classes to those calculated in \cite{Bravyi_Latone_Maslov_2022}.
In \Cref{tab:Sp2_gate_count,tab:Sp2_depth} we classify equivalence classes of symplectic matrices by circuit depth and two-qubit gate count.

\begin{table}[]
    \centering
    \begin{tabular}{| p{0.02\linewidth} | p{0.1\linewidth} | p{0.15\linewidth} |p{0.15\linewidth} |p{0.1\linewidth} |p{0.1\linewidth} |p{0.1\linewidth} |}
    \hline
n & $|\Sp(2n,2)|$  & Equiv Classes: This Work &  Equiv Classes: Bravyi et al & Max gate count & Max depth  \\
\hline
 1 & 6  & 1 & 1&0&0 \\
 2 & 720  & 2 & 4&2&1 \\
 3 & 1.45E+06  & 8 &27& 4& 4\\
 4 & 4.74E+10  & 109 &2,363& 6&4 \\
 5 & 2.48E+16  & 20,421 &4,322,659 & 9&5 \\
 6 & 2.08E+23  &  & 1.43E+11 &  & \\
\hline
    \end{tabular}
    \caption{Equivalence Classes of Symplectic Matrices. In this table, we compare the total number of binary symplectic matrices $|\Sp(2n,2)|$ to the equivalence classes used in our optimal algorithm and those used in \cite{Bravyi_Latone_Maslov_2022}. 
    We consider equivalence classes of symplectic matrices up to permutations of both columns and rows and  determine equivalence classes by mapping to a graph isomorphism problem using the method of \Cref{fig:sym_graph}.
    This significantly reduces the number of equivalence classes and computational complexity versus the method in \cite{Bravyi_Latone_Maslov_2022}.
    The maximum number of two-qubit gates required to generate any invertible matrix and the maximum depth required are noted in the final two columns.}
    \label{tab:Sp2_classes}
\end{table}
\subsubsection{Optimal Synthesis of Clifford Circuits}
The algorithm for optimal synthesis of Clifford circuits is as follows.
For a given symplectic matrix $A$, we first search the database for a symplectic matrix $B$ with the same graph isomorphism certificate. 
Using the canonical labelling, we determine a permutation $P_r$ on the row vertices and $P_c$ on column vertices which sends $G_B$ to $G_A$. 
Using the techniques in \cite{autQEC}, we map $P_r$ and $P_c$ to symplectic matrices $S_r$ and $S_c$ which correspond a combination of single-qubit Cliffords and qubit permutations such that $A = S_rBS_c$.
We return the stored circuit for $B$ prepended by the permutation and single-qubit Cliffords of $S_r$ and followed by those of $S_c$.
The gates of $S_c$ are then commuted through to the beginning of the circuit.

\subsection{A* Synthesis of Clifford Circuits}\label{sec:clifford_astar}
We now show how to extend the \CNOT A* synthesis algorithm to general Clifford circuits.
The key part of the algorithm is to choose a heuristic $h$ for the estimated number of two-qubit transvections required to reduce the symplectic matrix to a qubit permutation  followed by a set of single-qubit Cliffords.
\begin{added}
Our  heuristic uses the number of non-zero $F_{ij}$ matrices in each row and column of the symplectic matrix $A$ to estimate the number of transvections required to reduce $A$ to an identity matrix up to permutations and single-qubit Cliffords. 
These are tracked by using the matrix $\textit{R12}(A) := \textit{R1}(A) + \textit{R2}(A)$ where $\textit{R12}(A)_{ij} = 1$ if $F_{ij}$ is non-zero.
The column sum of $\textit{R12}(A) = n$ if and only if $A$ identity up to permutations and single-qubit Cliffords as in this case there are $n$ $F_{ij}$ of rank $2$.
The metrics we use for Clifford synthesis are direct analogies of the metrics of  \Cref{eq:H_heuristics} used for CNOT synthesis, but because $A^{-1} = \Omega A^T \Omega$ for a symplectic matrix $A$, we omit column sums of $\textit{R12}(A)$:
\end{added}
\begin{align}
H_\text{sum}(A) &:= (h_\text{sum}(\textit{R12}(A)) + h_\text{sum}(\textit{R12}(A^T)))/2n-1;\\
H_\text{prod}(A) &:= (h_\text{prod}(\textit{R12}(A)) + h_\text{prod}(\textit{R12}(A^T)))/2n.\label{eq:H12}
\end{align}
where $h_\text{sum}$ and $h_\text{prod}$ are as defined in \Cref{eq:H_heuristics}, but applied to the matrix $\textit{R12}(A)$ which indicates which $F_{ij}$ sub-matrices have rank 1 or 2 (see \Cref{eq:R12}).

Performing a regression analysis on the database of symplectic matrices generated as in \Cref{sec:clifford_optimal}, we find a stronger correlation for $H_\text{prod}$ versus $H_\text{sum}$.
When applied to the dataset of 1003 6-qubit symplectic matrices from \cite{Bravyi_Shaydulin_Hu_Maslov_2021}, we find that $rH_\text{prod}(A)$ gives a more robust optimisation which is very close to the known optimal 2-qubit gate count (\Cref{fig:Sp_astar_sensitivity}).
We found that the scalar heuristics of \Cref{eq:H_heuristics} did not perform well for general Clifford synthesis.

\begin{table}[]
    \centering
    \begin{tabular}{|l|r|r|r|r|r|}
    \hline
 $H_\text{sum}(A)$ & n=2 & 3 & 4 & 5 & 6 \\
 \hline
 R & 1.00 & 0.95 & 0.85 & 0.73 & 0.50 \\
 m & 1.00 & 1.61 & 1.53 & 1.69 & 0.98 \\
 b & 0.00 & -0.10 & 0.53 & 0.97 & 4.89 \\
 \hline
  \hline
$H_\text{prod}(A)$ & n=2 & 3 & 4 & 5 & 6 \\
 \hline
 R & 1.00 & 0.93 & 0.85 & 0.73 & 0.52 \\
 m & 1.44 & 2.98 & 3.93 & 6.44 & 4.84 \\
 b & 0.00 & -0.23 & -0.43 & -2.69 & 1.11 \\
 \hline
    \end{tabular}
    \caption{Correlation between optimal two-qubit gate-count and heuristics for binary symplectic $2n\times 2n$ matrix $A$. 
    The heuristics of \Cref{eq:H12} are based on the $\textit{R12}$ matrix which tracks which $F_{ij}$ have non-zero rank.
    We first calculate the correlation coefficient $R$, slope $m$ and intercept $b$ of the line of best fit for $H_\text{sum}(\textit{R12}(A))$ which is the sum of all entries in $\textit{R12}$. 
    We then do the same for $H_\text{prod}(\textit{R12}(A))$  which is the sum of the logs of the row and column sums of $\textit{R12}$.}
    \label{tab:correl_Sp}
\end{table}

\subsection{Clifford Synthesis Algorithm Benchmarks}
We  benchmarked our greedy, optimal and A* Clifford synthesis algorithms of \Cref{sec:clifford_greedy,sec:clifford_optimal,sec:clifford_astar} against existing  Clifford synthesis algorithms in the literature.

\subsubsection{Randomly Generated Cliffords on up to $n=64$ Qubits}
We benchmark all methods versus the Qiskit implementation of the Aaronson and Gottesman synthesis algorithm \cite{Aaronson_Gottesman_2004} and the results are in Figure \ref{fig:sym_benchmark}.
The following algorithms were used in the comparison:
\begin{itemize}
    \item The Qiskit greedy algorithm of \cite{Bravyi_Shaydulin_Hu_Maslov_2021} is a recursive algorithm which first calculates a circuit to disentangle one of the qubits from the others, then calls the algorithm on the remaining entangled qubits. 
    Note that this method operates quite differently to the greedy algorithm of \Cref{sec:clifford_greedy}.
    \item The pytket full peephole optimisation of \cite{Sivarajah_2021} is based on the peephole optimisation method set out in \cite{peephole} for reversible circuits. Peephole optimisation involves generating a set of known optimal circuits. 
    To optimise a circuit, the algorithm traverses small sections of the circuit and replaces them with an equivalent optimal circuit implementation.
    \item The rustiq package of \cite{GoubaultdeBrugiere2025graphstatebased} implements an graph-state based synthesis algorithm.
    \item The PyZX package of \cite{Kissinger_2020} synthesizes circuits using ZX calculus.
    \item The modified Volanto algorithm \cite{Volanto} which is described in \Cref{alg:volanto_synthesis}.
\end{itemize}

We used a dataset of 400 randomly generated symplectic matrices using the method of \cite{9605330}.
The results are summarized in \Cref{fig:sym_benchmark}.
We found that the A* algorithm matched optimal two-qubit gate counts up to $n=5$, and gave good results for our random dataset up to $n=16$ qubits.
To reduce the run-time of the A* algorithm we used a relatively small maximum queue size of 100. We used the heuristic scaling $r=3$ for the A* run.
For circuits on $n=16$ qubits, the greedy algorithm gave lower entangling gate counts than the A* algorithm and for $n>16$ qubits, it showed a clear separation against all other methods tested.
% For the greedy algorithm we found that appending the scalar heuristic $H_{sum}$ to the vector heuristic $\mathbf{h}$ improved  
A spreadsheet collating these results and giving the circuit implementations is available on our \href{https://github.com/m-webster/CliffordOpt/blob/main/paper_results/Sp_benchmark.xlsx}{GitHub repository}.

\subsubsection{Hamiltonian Dataset on up to $n=64$ Qubits}
\begin{figure}
     \centering
     \begin{subfigure}[t]{0.49\textwidth}
         \centering
    \includegraphics[width=\textwidth]{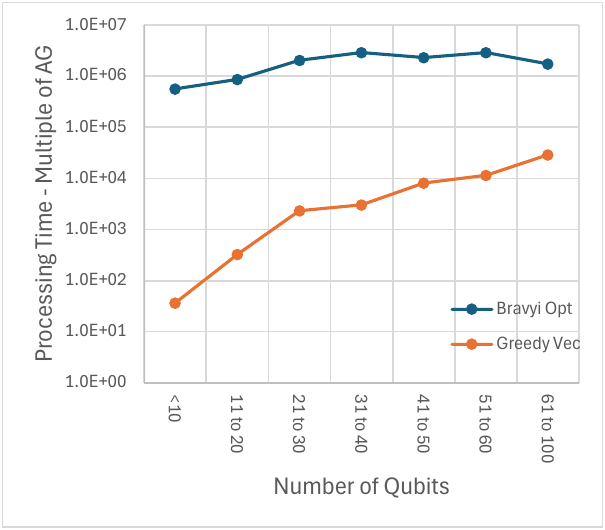}
    \caption{Processing Time }
     \end{subfigure}
     \hfill
     \begin{subfigure}[t]{0.49\textwidth}
         \centering
        \includegraphics[width=\textwidth]{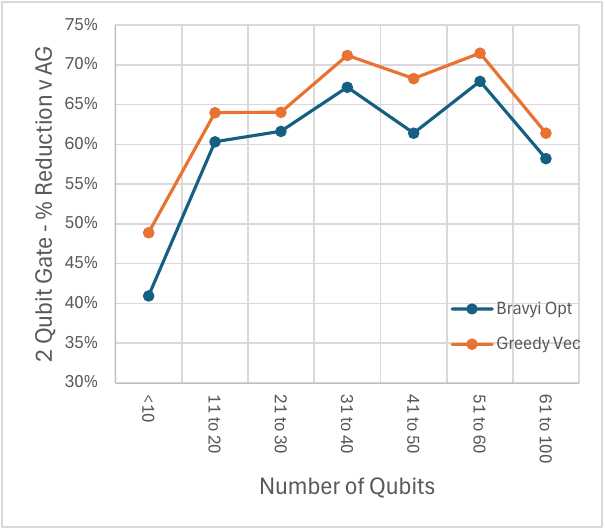}
        \caption{Two-Qubit Gate Count }
     \end{subfigure}
        \caption{Greedy Clifford synthesis benchmark - Hamiltonian dataset. \begin{added}We compare the processing time and two qubit gate count of the BSHM algorithm of \cite{Bravyi_Shaydulin_Hu_Maslov_2021} on Hamiltonian data sets. 
        The processing time is expressed as a multiple of the time taken by Aaronson and Gottesman (AG) synthesis (lower is better). Two-qubit gate count is in terms of the percentage reduction in gate count from AG synthesis (higher is better). Both are plotted against the number of qubits in the circuit. Note that we have batched these in ranges to make comparison easier - there are 2264 circuits in the data set.\end{added}}
        \label{fig:ham_benchmark}
\end{figure}

We then benchmarked using the Hamiltonian dataset of \cite{Bravyi_Shaydulin_Hu_Maslov_2021} - this is a series of circuits which represent the evolution of a Hamiltonian using various physical qubit layouts. 
The number of qubits in the circuits varied from 4 to 64. 
We used only the greedy Clifford synthesis, as optimal synthesis and A* were not possible on the larger circuits.
We benchmarked the greedy algorithm versus the BSHM algorithm in \cite{Bravyi_Shaydulin_Hu_Maslov_2021} followed by one round of the pytket FullPeepholeOptimise pass. 
This was done to ensure a fair comparison as the BSHM algorithm does not consider permutation equivalence classes of operators, but the pytket algorithm does. 
We found the greedy algorithm gave better results for both 2-qubit gate count and run time. 
The results are set out in \Cref{fig:ham_benchmark} and available on our \href{https://github.com/m-webster/CliffordOpt/blob/main/paper_results/bravyi_hamiltonian.xlsx}{GitHub repository}.

\subsubsection{Encoding Circuits}
Encoding circuits are an important class of circuits for Quantum Error Correction. Preparing the logical states $\ket{0}_L$ and $\ket{+}_L$ is used for instance in Knill \cite{KnillEC} and Steane \cite{SteaneEC} error correction, as well as in quantum algorithms. 

In \cite{zen2024} the authors use reinforcement learning to minimize the number of two qubit entangling gates in the $\ket{0}_L$ encoding circuit of different codes. They apply their technique on a number of codes, including non-CSS ones such as the perfect $[[5,1,3]]$ code. They also apply it to larger codes, such as the Golay $[[23,1,7]]$ code \cite{SteaneSimple}, where they find a $\ket{0}_L$ encoding circuit with $61$ entangling two qubit gates. In \cite{prgolay}, however, the authors found a circuit using $57$ \CNOT gates, by adapting Steane's Latin Rectangle method \cite{steane2004f} to minimize overlaps. Using our A* method, we found an encoding circuit using $56$ entangling two qubit gates, improving the best-known result. The results and comparisons are shown in Table \ref{tab:encoding_circuits}.

Furthermore, in \cite{peham2024, schmid2025} Boolean satisfiability solving techniques are leveraged to obtain optimal logical state preparation circuits for a number of CSS codes in terms of entangling gate count. 
% The authors obtain circuits aiming to minimize the entangling two qubit gate counts for $\ket{0}_L$ and $\ket{+}_L$ encoding. 
Applying the A* algorithm to the same codes gave circuits which match the optimal entangling gate counts. The greedy method gave results very close to optimal. 

The results and comparisons for logical state synthesis can be found \href{https://github.com/m-webster/CliffordOpt/blob/main/paper_results/RL_SAT_encoding.xlsx}{in our GitHub repository}.

\begin{table}[]
    \centering
    \begin{tabular}{|l|r|r|r|r|r|r|r|r|}
    \hline
Code&	RLFQC&	A* &	Greedy&	Pytket&	Qiskit&	Volanto&	PyZX&	Rustiq\\
\hline
$[[5,1,3]]$ Perfect Code&	6&	6&	6&	7&	8&	7&	10&	12\\
$[[7,1,3]]$ Steane Code&	8&	8&	8&	10&	11&	12&	9&	22\\
$[[9,1,3]]$ Shor Code&	8&	8&	8&	10&	10&	11&	9&	11\\
$[[15,1,3]]$ Reed-Muller&	22&	22&	22&	31&	32&	37&	24&	55\\
$[[17,1,5]]$ Colour Code&	37&	31&	33&	43&	43&	42&	42&	52\\
$[[19,1,5]]$ Colour Code&	49&	45&	49&	61&	61&	70&	67&	151\\
$[[23,1,7]]$ Golay Code&	61&	56&	69&	96&	97&	118&	138&	237\\
$[[25,1,5]]$ Surface Code&	54&	54&	53&	72&	72&	61&	71&	84\\
\hline
TOTAL&	245&	230&	248&	330&	334&	358&	370&	624\\
\hline
    \end{tabular}
    \caption{Entangling two qubit gate count in $\ket{0}_L$ encoding circuit for the codes in \cite{zen2024}. We compare the reinforcement learning method of \cite{zen2024} \begin{added}(denoted RLFQC in the table above)\end{added} to our greedy and A* algorithms, as well as various other circuit optimisation algorithms.}
    \label{tab:encoding_circuits}
\end{table}

\begin{added}
\subsubsection{Comparison with QSynth SAT Synthesis}
In reference \cite{shaikSATsynthesis} the authors introduce the qsynth SAT solver methods for optimising the two-qubit gate count and depth of Clifford circuits.
In \Cref{tab:SAT_Clifford}, we compare qsynth to our optimal, A* and greedy algorithms for random Clifford circuit on between 3 and 7 qubits. 
We see that our optimal and A* algorithms match the SAT solver two-qubit gate counts on circuits of up to 5 qubits with much lower processing time. 
On circuits of 6 qubits or more, the SAT methods timed out in almost all instances, whereas our A* and Greedy methods completed in all cases and had the same or lower two qubit gate count than the estimates in \cite{shaikSATsynthesis}.

\begin{table}[h!]												
\begin{center}												
\begin{tabular}{ |l|	r|	r|	r|	r|	r|	r|	r|	r|	r|	r|	r|}	\hline
&	\textbf{tket}&	\multicolumn{2}{|l|}{\textbf{qsynth-f}}&		\multicolumn{2}{|l|}{\textbf{qsynth-b}}&		\multicolumn{2}{|l|}{\textbf{Optimal}}&		\multicolumn{2}{|l|}{\textbf{A*}}&		\multicolumn{2}{|l|}{\textbf{Greedy}}\\		\hline
\textbf{Instance}&	\textbf{m}&	\textbf{m}&	\textbf{t}&	\textbf{m}&	\textbf{t}&	\textbf{m}&	\textbf{t}&	\textbf{m}&	\textbf{t}&	\textbf{m}&	\textbf{t}\\	\hline
3q05306&	2&	2&	1.0E-01&	2&	9.0E-02&	2&	1.1E-02&	2&	3.7E-01&	2&	3.0E-01\\	
3q33936&	3&	3&	7.0E-02&	3&	6.0E-02&	3&	1.4E-03&	3&	5.0E-03&	3&	6.4E-03\\	
3q50494&	3&	3&	6.0E-02&	3&	5.0E-02&	3&	1.4E-03&	3&	4.8E-03&	3&	6.2E-03\\	
3q55125&	3&	3&	6.0E-02&	3&	5.0E-02&	3&	1.3E-03&	3&	4.9E-03&	3&	6.4E-03\\	
3q99346&	3&	3&	6.0E-02&	3&	5.0E-02&	3&	1.1E-03&	3&	4.9E-03&	3&	6.3E-03\\	
4q05306&	6&	5&	4.4E-01&	5&	5.0E-01&	5&	2.8E-03&	5&	2.1E-01&	5&	1.9E-02\\	
4q33936&	8&	5&	3.5E-01&	5&	7.1E-01&	5&	1.2E-03&	5&	2.1E-01&	5&	1.6E-02\\	
4q50494&	4&	4&	1.0E-01&	4&	7.0E-02&	4&	1.2E-03&	4&	8.7E-03&	4&	1.0E-02\\	
4q55125&	6&	5&	4.3E-01&	5&	3.3E-01&	5&	1.1E-03&	5&	2.0E-01&	6&	2.0E-02\\	
4q99346&	6&	5&	4.3E-01&	5&	3.5E-01&	5&	1.2E-03&	5&	2.2E-01&	5&	1.7E-02\\	
5q05306&	11&	8&	3.5E+01&	8&	1.5E+02&	8&	4.9E-03&	8&	2.8E+00&	8&	3.6E-02\\	
5q33936&	10&	8&	3.4E+01&	8&	1.3E+02&	8&	3.5E-03&	8&	2.7E+00&	9&	4.7E-02\\	
5q50494&	11&	7&	2.2E+01&	7&	5.7E+01&	7&	3.0E-03&	7&	1.4E+00&	8&	3.6E-02\\	
5q55125&	9&	7&	4.5E+00&	7&	9.3E+00&	7&	2.3E-03&	7&	1.7E+00&	7&	2.8E-02\\	
5q99346&	10&	6&	1.5E+00&	6&	7.9E+00&	6&	2.3E-03&	6&	7.4E-01&	6&	2.7E-02\\	
6q05306&	15&	15&	TO&	11&	TO&	-&	-&	11&	6.9E+00&	12&	9.0E-02\\	
6q33936&	14&	14&	TO&	11&	TO&	-&	-&	10&	5.8E+00&	12&	9.9E-02\\	
6q50494&	14&	14&	TO&	12&	TO&	-&	-&	11&	6.6E+00&	11&	7.7E-02\\	
6q55125&	14&	10&	6.5E+03&	10&	TO&	-&	-&	10&	5.8E+00&	10&	7.6E-02\\	
6q99346&	15&	-&	TO&	11&	TO&	-&	-&	11&	6.8E+00&	11&	8.4E-02\\	
7q05306&	24&	-&	TO&	19&	TO&	-&	-&	15&	1.4E+01&	16&	1.6E-01\\	
7q33936&	20&	-&	TO&	17&	TO&	-&	-&	14&	1.4E+01&	16&	1.6E-01\\	
7q50494&	20&	-&	TO&	-&	TO&	-&	-&	14&	1.3E+01&	17&	1.8E-01\\	
7q55125&	18&	-&	TO&	17&	TO&	-&	-&	14&	1.3E+01&	14&	1.2E-01\\	
7q99346&	20&	-&	TO&	19&	TO&	-&	-&	14&	1.2E+01&	16&	1.4E-01\\	\hline
\end{tabular}		
\caption{QSynth SAT synthesis benchmark, random Clifford circuits. This table lists the number of two-qubit gates (columns labelled m) and processing time (t) for tket and the qsynth-f and qsynth-b from the `with permutation' section of Table 6 of \cite{shaikSATsynthesis}. We compare these with results for the Optimal, A* and Greedy algorithms of \Cref{sec:clifford_optimal,sec:clifford_astar,sec:clifford_greedy}. For A*, we use a queue size of 100. We see that the optimal algorithm completes synthesis for circuits up to 5 qubits with processing time several orders of magnitude less than qsynth. For circuits with 6 or more qubits, qsynth timed out in all but one instance - for these cases,  TO is indicated in the processing time column and an upper bound is given where possible. The A* and greedy methods gave the same or lower two-qubit gate count with significantly lower processing time for all circuits in the data set.}
    \label{tab:SAT_Clifford}
\end{center}
\end{table}												

\end{added}

\section{Conclusion and Open Questions}
In this work, we have set out algorithms for synthesis of \CNOT and general Clifford circuits which have lower two-qubit gate count than existing methods, and can also be applied to optimise for circuit depth.

The optimal synthesis algorithms can be used for \CNOT circuits on $n\le 7$ qubits and Clifford circuits on $n \le 5$ qubits. 
By using the optimised coding techniques of \cite{Bravyi_Latone_Maslov_2022} it may well be possible to extend this size range.
Determining a closed expression for the maximum circuit depth and two-qubit gate-count to reach any desired \CNOT or Clifford circuit would be an interesting question to address.

The A* synthesis algorithms closely matches optimal 2-qubit gate counts where these are available, and can be used for intermediate-size circuits. The algorithm has exponential run time in the worst case, but this can be managed by shortening the maximum priority queue length. 
Using the data from optimal synthesis, we could use more sophisticated machine learning methods to determine the heuristic $h$ and this has potential to improve the accuracy of the method.

The greedy algorithm out-performs existing methods for larger circuit sizes.
The output of the greedy algorithm could be used as input to pattern-matching algorithms such as \cite{Bravyi_Shaydulin_Hu_Maslov_2021,Sivarajah_2021}.

We have implemented our algorithms in python to enable rapid development and maximal re-use of code between the algorithms.
The speed of the algorithms could be improved by porting them to C or Rust, allowing them to be used effectively on larger circuits.
The algorithms could also be modified to take into account the gate set and connectivity of particular devices, though we have not built this functionality.

% These algorithms have been made available to the classical and quantum computing communities in our \href{https://github.com/m-webster/CliffordOpt}{GitHub repository} \cite{webster_m-cliffordopt}. 
The software developed in this work and output data underpinning our results is available in a \href{https://github.com/m-webster/CliffordOpt}{python GitHub repository} and \href{https://doi.org/10.5281/zenodo.21904613}{Zenodo}.

\section{Acknowledgments}
 MW and DEB are supported by the Engineering and Physical Sciences Research Council [grant number EP/W032635/1 and EP/S005021/1].
SK and DEB are supported by the Engineering and Physical Sciences Research Council [grant number EP/Y004620/1 and EP/T001062/1]. The authors acknowledge the use of the UCL Kathleen High Performance Computing Facility (Kathleen@UCL), and associated support services, in the completion of this work. The authors would like to thank Hasan Sayginel, Nicholas Fazio and Zhenyu Cai for helpful discussions. 

\section{Author Contribution Statement}
MW conceived of and designed the study, including the optimal, A*, and greedy synthesis algorithms. 
SK surveyed the literature and contributed insights to the algorithm design and benchmarking. 
DEB supervised the project and secured funding. 
MW wrote the original draft of the manuscript. 
All authors reviewed and edited the manuscript. 
No AI or Large Language Models were used in producing this work.

\medskip

\bibliographystyle{plainnat}
\bibliography{references}

\begin{thebibliography}{45}
\providecommand{\natexlab}[1]{#1}
\providecommand{\url}[1]{\texttt{#1}}
\expandafter\ifx\csname urlstyle\endcsname\relax
  \providecommand{\doi}[1]{doi: #1}\else
  \providecommand{\doi}{doi: \begingroup \urlstyle{rm}\Url}\fi

\bibitem[Aaronson and Gottesman(2004)]{Aaronson_Gottesman_2004}
Scott Aaronson and Daniel Gottesman.
\newblock Improved simulation of stabilizer circuits.
\newblock \emph{Physical Review A}, 70\penalty0 (5):\penalty0 052328, November 2004.
\newblock ISSN 1050-2947, 1094-1622.
\newblock \doi{10.1103/PhysRevA.70.052328}.

\bibitem[Baart et~al.(2015)]{Baart2015SinglespinC}
Timothy~A. Baart et~al.
\newblock Single-spin {CCD}.
\newblock \emph{Nature nanotechnology}, 11 4:\penalty0 330--4, 2015.
\newblock \doi{10.1038/nnano.2015.291}.

\bibitem[Bravyi et~al.(2021)Bravyi, Shaydulin, Hu, and Maslov]{Bravyi_Shaydulin_Hu_Maslov_2021}
Sergey Bravyi, Ruslan Shaydulin, Shaohan Hu, and Dmitri Maslov.
\newblock {Clifford} circuit optimization with templates and symbolic {P}auli gates.
\newblock \emph{Quantum}, 5:\penalty0 580, November 2021.
\newblock ISSN 2521-327X.
\newblock \doi{10.22331/q-2021-11-16-580}.

\bibitem[Bravyi et~al.(2022)Bravyi, Latone, and Maslov]{Bravyi_Latone_Maslov_2022}
Sergey Bravyi, Joseph~A. Latone, and Dmitri Maslov.
\newblock 6-qubit optimal {C}lifford circuits.
\newblock \emph{npj Quantum Information}, 8\penalty0 (1):\penalty0 79, July 2022.
\newblock ISSN 2056-6387.
\newblock \doi{10.1038/s41534-022-00583-7}.

\bibitem[Christensen et~al.(2025)Christensen, J\o{}rgensen, Pavlogiannis, and van~de Pol]{christensen2025exactsizesminimalcnot}
Jens~Emil Christensen, S\o{}ren~Fuglede J\o{}rgensen, Andreas Pavlogiannis, and Jaco van~de Pol.
\newblock On exact sizes of minimal {CNOT} circuits.
\newblock In \emph{Reversible Computation: 17th International Conference, RC 2025, Odense, Denmark, July 3–4, 2025, Proceedings}, page 71–88, Berlin, Heidelberg, 2025. Springer-Verlag.
\newblock ISBN 978-3-031-97062-7.
\newblock \doi{10.1007/978-3-031-97063-4_6}.

\bibitem[Cirac and Zoller(1995)]{trapped_ions}
J.~I. Cirac and P.~Zoller.
\newblock Quantum computations with cold trapped ions.
\newblock \emph{Phys. Rev. Lett.}, 74:\penalty0 4091--4094, May 1995.
\newblock \doi{10.1103/PhysRevLett.74.4091}.

\bibitem[De~Brugi\`{e}re et~al.(2021)]{Brugiere1}
Timoth\'{e}e~Goubault De~Brugi\`{e}re et~al.
\newblock Gaussian elimination versus greedy methods for the synthesis of linear reversible circuits.
\newblock \emph{ACM Transactions on Quantum Computing}, 2\penalty0 (3), September 2021.
\newblock \doi{10.1145/3474226}.

\bibitem[Fazio et~al.(2025)Fazio, Webster, and Cai]{fazio2025lowoverheadmagicstatecircuits}
Nicholas Fazio, Mark Webster, and Zhenyu Cai.
\newblock Low-overhead magic state circuits with transversal {{CNOT}}s.
\newblock 2025.
\newblock \doi{10.48550/arXiv.2501.10291}.

\bibitem[Goubault~de Brugi{\`{e}}re et~al.(2025)Goubault~de Brugi{\`{e}}re, Martiel, and Vuillot]{GoubaultdeBrugiere2025graphstatebased}
Timoth{\'{e}}e Goubault~de Brugi{\`{e}}re, Simon Martiel, and Christophe Vuillot.
\newblock A graph-state based synthesis framework for {C}lifford isometries.
\newblock \emph{{Quantum}}, 9:\penalty0 1589, January 2025.
\newblock ISSN 2521-327X.
\newblock \doi{10.22331/q-2025-01-14-1589}.

\bibitem[Hart et~al.(1968)Hart, Nilsson, and Raphael]{hart_nilsson_astar}
Peter~E. Hart, Nils~J. Nilsson, and Bertram Raphael.
\newblock A formal basis for the heuristic determination of minimum cost paths.
\newblock \emph{IEEE Transactions on Systems Science and Cybernetics}, 4\penalty0 (2):\penalty0 100--107, 1968.
\newblock \doi{10.1109/TSSC.1968.300136}.

\bibitem[Junttila and Kaski(2007)]{Bliss2}
Tommi Junttila and Petteri Kaski.
\newblock Engineering an efficient canonical labeling tool for large and sparse graphs.
\newblock In David Applegate et~al., editors, \emph{Proceedings of the Ninth Workshop on Algorithm Engineering and Experiments and the Fourth Workshop on Analytic Algorithms and Combinatorics}, pages 135--149. SIAM, 2007.
\newblock \doi{10.1137/1.9781611972870.13}.

\bibitem[Junttila and Kaski(2011)]{Bliss1}
Tommi Junttila and Petteri Kaski.
\newblock Conflict propagation and component recursion for canonical labeling.
\newblock In Alberto Marchetti{-}Spaccamela and Michael Segal, editors, \emph{Theory and Practice of Algorithms in (Computer) Systems -- First International {ICST} Conference, {TAPAS} 2011, Rome, Italy, April 18--20, 2011. Proceedings}, volume 6595 of \emph{Lecture Notes in Computer Science}, pages 151--162. Springer, 2011.
\newblock \doi{10.1007/978-3-642-19754-3\_16}.

\bibitem[Kielpinski et~al.(2002)Kielpinski, Monroe, and Wineland]{2002_Kielpinski}
David Kielpinski, C.R. Monroe, and D.J. Wineland.
\newblock Architecture for a large-scale ion-trap quantum computer.
\newblock \emph{Nature}, 417:\penalty0 709--11, 07 2002.
\newblock \doi{10.1038/nature00784}.

\bibitem[Kissinger and van~de Wetering(2020)]{Kissinger_2020}
Aleks Kissinger and John van~de Wetering.
\newblock {PyZX}: Large scale automated diagrammatic reasoning.
\newblock \emph{Electronic Proceedings in Theoretical Computer Science}, 318:\penalty0 229–241, May 2020.
\newblock ISSN 2075-2180.
\newblock \doi{10.4204/eptcs.318.14}.
\newblock URL \url{http://dx.doi.org/10.4204/EPTCS.318.14}.

\bibitem[Knill(2005)]{KnillEC}
E.~Knill.
\newblock Quantum computing with realistically noisy devices.
\newblock \emph{Nature}, 434\penalty0 (7029):\penalty0 39--44, March 2005.
\newblock ISSN 1476-4687.
\newblock \doi{10.1038/nature03350}.

\bibitem[Koenig and Smolin(2014)]{Koenig_Smolin_2014}
Robert Koenig and John~A. Smolin.
\newblock How to efficiently select an arbitrary {Clifford} group element.
\newblock \emph{Journal of Mathematical Physics}, 55\penalty0 (12):\penalty0 122202, December 2014.
\newblock ISSN 0022-2488, 1089-7658.
\newblock \doi{10.1063/1.4903507}.

\bibitem[Litinski(2019)]{Litinski_2019}
Daniel Litinski.
\newblock A game of surface codes: Large-scale quantum computing with lattice surgery.
\newblock \emph{Quantum}, 3:\penalty0 128, March 2019.
\newblock ISSN 2521-327X.
\newblock \doi{10.22331/q-2019-03-05-128}.

\bibitem[McKay and Piperno(2014)]{MCKAY201494}
Brendan~D. McKay and Adolfo Piperno.
\newblock Practical graph isomorphism, ii.
\newblock \emph{Journal of Symbolic Computation}, 60:\penalty0 94--112, 2014.
\newblock ISSN 0747-7171.
\newblock \doi{10.1016/j.jsc.2013.09.003}.

\bibitem[Meunier et~al.(2011)Meunier, Calado, and Vandersypen]{quantum_dots}
Tristan Meunier, Victor~E. Calado, and Lieven M.~K. Vandersypen.
\newblock Efficient controlled-phase gate for single-spin qubits in quantum dots.
\newblock \emph{Phys. Rev. B}, 83:\penalty0 121403, Mar 2011.
\newblock \doi{10.1103/PhysRevB.83.121403}.

\bibitem[Murphy and Kissinger(2023)]{Murphy_Kissinger_2023}
Ewan Murphy and Aleks Kissinger.
\newblock Global synthesis of {{CNOT}} circuits with holes.
\newblock \emph{Electronic Proceedings in Theoretical Computer Science}, 384:\penalty0 75–88, August 2023.
\newblock ISSN 2075-2180.
\newblock \doi{10.4204/EPTCS.384.5}.

\bibitem[O'Meara(1978)]{o1978symplectic}
O.T. O'Meara.
\newblock \emph{Symplectic Groups}.
\newblock Mathematical Surveys and Monographs. American Mathematical Society, 1978.
\newblock ISBN 9780821815168.
\newblock \doi{10.1090/surv/016}.

\bibitem[Paetznick and Reichardt(2012)]{prgolay}
Adam Paetznick and Ben~W. Reichardt.
\newblock Fault-tolerant ancilla preparation and noise threshold lower boudds for the 23-qubit golay code.
\newblock \emph{Quantum Info. Comput.}, 12\penalty0 (11–12):\penalty0 1034–1080, November 2012.
\newblock ISSN 1533-7146.
\newblock \doi{10.48550/arXiv.1106.2190}.

\bibitem[Patel et~al.(2008)Patel, Markov, and Hayes]{Patel_Markov_Hayes_2008}
K.N. Patel, I.L. Markov, and J.P. Hayes.
\newblock Optimal synthesis of linear reversible circuits.
\newblock \emph{Quantum Information and Computation}, 8\penalty0 (3 \& 4):\penalty0 282–294, March 2008.
\newblock ISSN 15337146, 15337146.
\newblock \doi{10.26421/QIC8.3-4-4}.

\bibitem[Peham et~al.(2025)Peham, Schmid, Berent, M\"uller, and Wille]{peham2024}
Tom Peham, Ludwig Schmid, Lucas Berent, Markus M\"uller, and Robert Wille.
\newblock Automated synthesis of fault-tolerant state preparation circuits for quantum error-correction codes.
\newblock \emph{PRX Quantum}, 6:\penalty0 020330, May 2025.
\newblock \doi{10.1103/PRXQuantum.6.020330}.

\bibitem[Pllaha et~al.(2021)Pllaha, Volanto, and Tirkkonen]{Pllaha_Volanto_Tirkkonen_2021}
Tefjol Pllaha, Kalle Volanto, and Olav Tirkkonen.
\newblock Decomposition of {Clifford} gates.
\newblock In \emph{2021 IEEE Global Communications Conference (GLOBECOM)}, page 01–06, December 2021.
\newblock \doi{10.1109/GLOBECOM46510.2021.9685501}.

\bibitem[Prasad et~al.(2006)Prasad, Shende, Markov, Hayes, and Patel]{peephole}
Aditya~K. Prasad, Vivek~V. Shende, Igor~L. Markov, John~P. Hayes, and Ketan~N. Patel.
\newblock Data structures and algorithms for simplifying reversible circuits.
\newblock \emph{J. Emerg. Technol. Comput. Syst.}, 2\penalty0 (4):\penalty0 277–293, October 2006.
\newblock ISSN 1550-4832.
\newblock \doi{10.1145/1216396.1216399}.

\bibitem[Rengaswamy et~al.(2018)Rengaswamy, Calderbank, Pfister, and Kadhe]{Rengaswamy_Calderbank_Pfister_Kadhe_2018}
Narayanan Rengaswamy, Robert Calderbank, Henry~D. Pfister, and Swanand Kadhe.
\newblock Synthesis of logical {Clifford} operators via symplectic geometry.
\newblock In \emph{2018 IEEE International Symposium on Information Theory (ISIT)}, page 791–795, Vail, CO, USA, June 2018. IEEE.
\newblock ISBN 978-1-5386-4781-3.
\newblock \doi{10.1109/ISIT.2018.8437652}.

\bibitem[Rodriguez et~al.(2025)Rodriguez, Robinson, Jepsen, et~al.]{rodriguez2024experimentaldemonstrationlogicalmagic}
Pedro~Sales Rodriguez, John~M. Robinson, Paul~Niklas Jepsen, et~al.
\newblock Experimental demonstration of logical magic state distillation.
\newblock \emph{Nature}, 645:\penalty0 620–625, Sep 2025.
\newblock \doi{10.1038/s41586-025-09367-3}.

\bibitem[Sayginel et~al.(2025)Sayginel, Koutsioumpas, Webster, Rajput, and Browne]{autQEC}
Hasan Sayginel, Stergios Koutsioumpas, Mark Webster, Abhishek Rajput, and Dan~E. Browne.
\newblock Fault-tolerant logical clifford gates from code automorphisms.
\newblock \emph{PRX Quantum}, 6:\penalty0 030343, Sep 2025.
\newblock \doi{10.1103/vf7v-cpq9}.

\bibitem[Schaeffer and Perkowski(2014)]{schaeffer2014costminimizationapproachsynthesis}
Ben Schaeffer and Marek Perkowski.
\newblock A cost minimization approach to synthesis of linear reversible circuits.
\newblock 2014.
\newblock \doi{10.48550/arXiv.1407.0070}.

\bibitem[Schmid et~al.(2025)Schmid, Peham, Berent, Müller, and Wille]{schmid2025}
Ludwig Schmid, Tom Peham, Lucas Berent, Markus Müller, and Robert Wille.
\newblock Deterministic fault-tolerant state preparation for near-term quantum error correction: Automatic synthesis using boolean satisfiability.
\newblock 2025.
\newblock \doi{10.48550/arXiv.2501.05527}.

\bibitem[Shaik and van~de Pol(2025)]{shaikSATsynthesis}
Irfansha Shaik and Jaco van~de Pol.
\newblock {CNOT-Optimal Clifford Synthesis as SAT}.
\newblock In Jeremias Berg and Jakob Nordstr\"{o}m, editors, \emph{28th International Conference on Theory and Applications of Satisfiability Testing (SAT 2025)}, volume 341 of \emph{Leibniz International Proceedings in Informatics (LIPIcs)}, pages 28:1--28:21, Dagstuhl, Germany, 2025. Schloss Dagstuhl -- Leibniz-Zentrum f{\"u}r Informatik.
\newblock ISBN 978-3-95977-381-2.
\newblock \doi{10.4230/LIPIcs.SAT.2025.28}.

\bibitem[Simmons(2024)]{Photonic}
Stephanie Simmons.
\newblock Scalable fault-tolerant quantum technologies with silicon color centers.
\newblock \emph{PRX Quantum}, 5:\penalty0 010102, Mar 2024.
\newblock \doi{10.1103/PRXQuantum.5.010102}.

\bibitem[Sivarajah et~al.(2020)]{Sivarajah_2021}
Seyon Sivarajah et~al.
\newblock t|ket⟩: a retargetable compiler for {NISQ} devices.
\newblock \emph{Quantum Science and Technology}, 6\penalty0 (1):\penalty0 014003, nov 2020.
\newblock \doi{10.1088/2058-9565/ab8e92}.

\bibitem[Steane(1996)]{SteaneSimple}
A.~M. Steane.
\newblock Simple quantum error-correcting codes.
\newblock \emph{Phys. Rev. A}, 54:\penalty0 4741--4751, Dec 1996.
\newblock \doi{10.1103/PhysRevA.54.4741}.

\bibitem[Steane(1997)]{SteaneEC}
A.~M. Steane.
\newblock Active stabilization, quantum computation, and quantum state synthesis.
\newblock \emph{Phys. Rev. Lett.}, 78:\penalty0 2252--2255, Mar 1997.
\newblock \doi{10.1103/PhysRevLett.78.2252}.

\bibitem[Steane(2004)]{steane2004f}
Andrew~M. Steane.
\newblock Fast fault-tolerant filtering of quantum codewords.
\newblock 2004.
\newblock \doi{10.48550/arXiv.quant-ph/0202036}.

\bibitem[Stromberg()]{stromberg_treap}
Daniel~R Stromberg.
\newblock treap: {Python} implementation of treaps.
\newblock URL \url{http://stromberg.dnsalias.org/~dstromberg/treap/}.

\bibitem[Van Den~Berg(2021)]{9605330}
Ewout Van Den~Berg.
\newblock A simple method for sampling random {C}lifford operators.
\newblock In \emph{2021 IEEE International Conference on Quantum Computing and Engineering (QCE)}, pages 54--59, 2021.
\newblock \doi{10.1109/QCE52317.2021.00021}.

\bibitem[Vandersypen and Chuang(2005)]{NMR}
Lieven M.~K. Vandersypen and Isaac~L. Chuang.
\newblock {NMR} techniques for quantum control and computation.
\newblock \emph{Rev. Mod. Phys.}, 76:\penalty0 1037--1069, Jan 2005.
\newblock \doi{10.1103/RevModPhys.76.1037}.

\bibitem[Volanto(2023)]{Volanto}
Kalle Volanto.
\newblock Minimizing the number of two-qubit gates in {C}lifford circuits.
\newblock Master's thesis, Aalto University, March 2023.
\newblock URL \url{https://urn.fi/URN:NBN:fi:aalto-202303262589}.

\bibitem[Webster(2025)]{webster_m-cliffordopt}
Mark Webster.
\newblock {CliffordOpt}: {Optimisation} of {Clifford} {Circuits}, May 2025.
\newblock URL \url{https://doi.org/10.5281/zenodo.21904614}.

\bibitem[Xiao et~al.(2024)]{2024_rydberg_atoms}
Yang Xiao et~al.
\newblock Effective nonadiabatic holonomic swap gate with {R}ydberg atoms using invariant-based reverse engineering.
\newblock \emph{Phys. Rev. A}, 109:\penalty0 062610, Jun 2024.
\newblock \doi{10.1103/PhysRevA.109.062610}.

\bibitem[Zen et~al.(2025)Zen, Olle, Colmenarez, Puviani, M\"uller, and Marquardt]{zen2024}
Remmy Zen, Jan Olle, Luis Colmenarez, Matteo Puviani, Markus M\"uller, and Florian Marquardt.
\newblock Quantum circuit discovery for fault-tolerant logical state preparation with reinforcement learning.
\newblock \emph{Phys. Rev. X}, 15:\penalty0 041012, Oct 2025.
\newblock \doi{10.1103/gqpr-dgz7}.

\bibitem[Živković(2006)]{ZIVKOVIC2006310}
Miodrag Živković.
\newblock Classification of small (0,1) matrices.
\newblock \emph{Linear Algebra and its Applications}, 414\penalty0 (1):\penalty0 310--346, 2006.
\newblock ISSN 0024-3795.
\newblock \doi{10.1016/j.laa.2005.10.010}.

\end{thebibliography}

\appendix

\section{Algorithm Pseudocode}
In this appendix, we present pseudocode for the algorithms used in this paper. 
We start by describing  Gaussian elimination and Volanto synthesis for \CNOT and general Clifford circuits respectively, highlighting their similarities.
We then introduce our greedy algorithm with its modifications to allow for early termination when trapped in local minima and optimisation of circuit depth rather than gate-count.
Finally, we describe the A* algorithm and the closely related optimal synthesis algorithm.

\subsection{Modified Gaussian Elimination}
We first present a modified version of Gaussian elimination which can be used for \CNOT circuit synthesis.
The modified algorithm has the following variations to the usual one.
Firstly, it scans down rows and across columns to perform eliminations via column operations.
Secondly, it reduces invertible matrices to a permutation matrix $P$ rather than to identity.
To assist with this, it maintains a set \texttt{pOptions} of unused pivots.
% If the matrix is full rank, the algorithm finds a pivot \texttt{p} for each column i such that $A_{pi}=1$; if not, the algorithm returns \texttt{FALSE}.
The output of the algorithm is the permutation matrix \texttt{P} plus a \begin{added}sequence\end{added} of \CNOT gates \texttt{opList} which when applied to \texttt{P} give \texttt{A}.

\begin{algorithm}[H]
\caption{Modified Gaussian Elimination}\label{alg:gaussian}
\end{algorithm}
\vspace{-4ex}
\begin{algorithmic}
\State\textbf{Input:} 
\State \texttt{A}: a $n\times n$ binary invertible matrix
\State\textbf{Output:}  
\State \texttt{P}: a permutation matrix; AND
\State \texttt{opList}: series of \CNOT gates yielding \texttt{A} when applied to \texttt{P}; OR 
\State \texttt{FALSE}: if \texttt{A} is not invertible
\State\textbf{Method:} 
\State \texttt{opList := [], pOptions := [0..n-1], P := copy(A)}
\For{\texttt{i in [0..n-1]}}\Comment{i is the current row}
\State \texttt{R := [p for p in pOptions if P[i,p]=1] }\Comment{unused pivots which are non-zero in row i}
\If{\texttt{len(R)=0}}
\State \texttt{return FALSE}\Comment{the matrix is not full rank}
\EndIf
\State \texttt{p := R.removeMin()} \Comment{remove smallest element of \texttt{R}, save as \texttt{p}}
\State \texttt{pOptions.remove(p)} \Comment{remove \texttt{p} from unused pivots}
\For{\texttt{j in R}}
\State \texttt{op:= (\CNOT,(p,j))}
\State \texttt{applyOp(P,op), opList.append(op)}
\EndFor
\EndFor
\State \texttt{return P, inverse(opList)}
\end{algorithmic}

\subsection{Symplectic Representation of Transvections}\label{sec:transvection_proof}
In this section, we find the symplectic matrix representation of the transvections defined in \Cref{eq:transvection_def}.

\paragraph{Proposition}
Let $T_P:=\exp\left(\frac{i\pi}{4}(I - P)\right)$ where $P$ is a Pauli with binary representation $\mathbf{v}$ satisfying $P^2=I$.
$T_P$ has symplectic representation $I + \Omega\mathbf{v}^T\mathbf{v}$

\paragraph{Proof}
Define the projectors $A_\pm:=\frac{1}{2}(I \pm P)$.
Note that $A_\pm^2 = A_\pm$ and $I = A_+ + A_-$.
Let $A_-(\theta) := \exp\left({i\theta}A_-\right)$ so that $T_P = A_-(\frac{\pi}{2})$. 
Expanding $A_-(\theta)$, we have:
\begin{align*}
    A_-(\theta) &:= \exp\left({i\theta}A_-\right) \\&= I + \frac{i\theta}{1!}A_- -  \frac{\theta^2}{2!}A_--i\frac{\theta^3}{3!}A_-+\frac{\theta^4}{4!}A_- + ...\\&=A_+ + A_-(\cos \theta + i \sin \theta )
\end{align*}
Setting $\theta := \frac{\pi}{2}$ and $\theta := \pi$:
\begin{align*}
T_P &= A_-\left(\frac{\pi}{2}\right) =A_+ + i A_-,\\
T_P^2 &=  A_-(\pi) = A_+ -A_- = \frac{1}{2}(I + P - (I - P)) = P.
\end{align*}
Since $P^2 = I$,  $T_PT_P^3 = I$ and so $T_P^{-1} = PT_P$.
Now consider the action of $T_P$ on a Pauli operator $Q$ under conjugation. If $P$ and $Q$ commute then $QP=PQ$ and $QA_\pm = A_\pm A$ so that:
\begin{added}
\begin{align*}
T_P Q T_P^{-1} &= T_PQP(A_++iA_-) = T_PP(A_++iA_-)Q = T_P T_P^{-1} Q = Q.
\end{align*}
\end{added}
Otherwise, if $P$ and $Q$ anti-commute:
\begin{align*}
T_P Q T_P^{-1} &= (A_+ + iA_-)QP(A_++iA_-) \\&= -(A_+ + iA_-)P(A_-+iA_+)Q \\&= -\frac{1}{i}(A_+ + iA_-)(A_+-iA_-)PQ \\&= i(A_+ + A_-)PQ = iPQ.
\end{align*}
We claim the symplectic matrix $I + \Omega\mathbf{v}^T\mathbf{v}$ has the same action acting by right matrix multiplication on vector representations of Paulis. 
Let $Q$ have vector representation $\mathbf{u}$. 
The expression $\mathbf{u} \Omega\mathbf{v}^T$  is the symplectic inner product and is zero if $Q$ commutes with $P$ and 1 otherwise.  
If $P$ and $Q$ commute then
$\mathbf{u}(I + \Omega\mathbf{v}^T\mathbf{v}) = \mathbf{u} +  (\mathbf{u} \Omega\mathbf{v}^T)\mathbf{v} = \mathbf{u}$ and so $Q$ is mapped to itself.
If $P$ and $Q$ anti-commute then
$\mathbf{u}(I + \Omega\mathbf{v}^T\mathbf{v}) = \mathbf{u} +  (\mathbf{u} \Omega\mathbf{v}^T)\mathbf{v} = \mathbf{u} + \mathbf{v}$. 
Because  addition of vector representations corresponds to multiplication of Paulis up to $\pm 1$, $Q$ is mapped to $PQ$ up to a global phase.

\subsection{Modified Volanto Synthesis Algorithm}
In this section, we present a modified version of Volanto Clifford synthesis (see \cite{Volanto}).
The algorithm works by reducing the symplectic matrix representation of a Clifford circuit to a matrix which represents a qubit permutation followed by a series of single-qubit Clifford gates (see \Cref{sec:clifford_greedy}).  
The modified Volanto algorithm works by scanning down the rows of the matrix.
For each row, it ensures that there is exactly one $F_{ij}$ matrix of rank $2$ and none of rank $1$ (see \Cref{eq:Fij}).
The algorithm does this by reducing pairs of rank $2$ $F_{ij}$ in the row to rank $1$, then reduces the remaining rank $1$ $F_{ij}$ to all-zero matrices. 

In the tableau view of symplectic matrices \cite{Aaronson_Gottesman_2004},  $F_{ij}$ represents a pair of single-qubit Paulis acting on qubit $j$ - one of which is part of a stabiliser in row $i$ and one of which is an anti-commuting destabiliser in row $i+n$. 
It can easily be verified that single-qubit Paulis acting on qubit $j$ anti-commute if and only if $F_{ij}$ is of rank $2$. 
Hence there are always an odd number of full-rank $F_{ij}$ in any row $i$ or column $j$.

\begin{added}
In the material below, we use a standardised transvection form: $$\text{Tv}(a,b,c,d)_{jk} := \sqrt{i^{ac+bd}X_j^a Z_j^b X_k^c Z_k^d}$$ where $a,b,c$ and $d$ are either zero or one and $j,k$ indicate the two qubits the transvection acts on.
To eliminate pairs $F_{ij}, F_{ik}$ in row $i$ which are both of rank $2$, we apply the transvection $\textit{T2}_{ijk} =\text{Tv}(a,b,c,d)_{jk}$ on the right where $(a,b)$ is the first row of $F_{ij}$ and $(c,d)$ is the second row of $F_{ik}$.
Given that there is an odd number of $F_{ij}$ of rank 2 in each row, this leaves a single $F_{ij}$ of rank 2 in the row and we use column $j$ as the pivot for row $i$.

We eliminate $F_{ik}$ of rank 1 as follows. The pivot matrix $F_{ij}$ is of rank 2 so it is invertible by assumption.
As $F_{ik}$ of rank 1, $F_{ij}^{-1} F_{jk} = \begin{bmatrix}b\\a\end{bmatrix} \begin{bmatrix}c&d\end{bmatrix}$ where $a := 1$ if the second  row of $F_{ij}^{-1} F_{jk}$ is non-zero and $b:=1$ if the first row is non-zero. We set  $c  := 1$ if the first column is non-zero and $d:=1$ if the second column is non-zero.
Applying the transvection $\textit{T1}_{ijk}:=\text{Tv}(a,b,c,d)_{jk}$ on the right reduces $F_{ik}$ to the all-zero matrix.  
\end{added}

The result of the algorithm is a symplectic matrix which has exactly one matrix $F_{ij}$ of rank 2 for each row $i$ and column $j$, and no matrices of rank $1$. 
This can be interpreted as a qubit permutation which is defined by the pivot for each row followed by a series of single-qubit Clifford gates. 
The modified algorithm returns the single qubit Clifford gates, a qubit permutation and the reversed list of 2-qubit transvections.

Note the similarities to the modified Gaussian elimination algorithm of \Cref{alg:gaussian}.
In particular it maintains a list of unused pivots and returns \texttt{FALSE} if A is not symplectic.

\begin{algorithm}[H]
\caption{Modified Volanto Synthesis}\label{alg:volanto_synthesis}
\end{algorithm}
% \end{figure}
\vspace{-4ex}
\begin{algorithmic}
\State\textbf{Input:} 
\State \texttt{A}: a $2n\times 2n$ binary symplectic matrix
\State\textbf{Output:}  
\State \texttt{P}: the final reduced matrix; AND
\State{\texttt{opList}: series of 2-qubit transvections yielding \texttt{A} when applied to \texttt{P}; OR}
\State{\texttt{FALSE}: if \texttt{A} is not symplectic}
\State\textbf{Method:} 
\State \texttt{opList := [], pOptions := [0..n-1], P := copy(A)}
\For{\texttt{i in [0..n-1]}}\Comment{\texttt{i} is the current row}
\State \texttt{R := [p for p in pOptions if Rank(F(P,i,p) = 2]} \Comment{unused pivots with Rank($F_{ip}$)=2}
\If{\texttt{len(R) mod 2 = 0}}
\State return \texttt{FALSE}\Comment{symplectic matrices have an odd number of $F_{ip}$ of rank 2}
\EndIf
\State \texttt{p := R.removeMin()} \Comment{remove smallest element of \texttt{R}, save as \texttt{p}}
\State \texttt{pOptions.remove(p}) \Comment{remove \texttt{p} from unused pivots}
\For{\texttt{a in [0..len(R)/2 - 1]}}
\State \texttt{j:=R[2a], k:=R[2a+1]}
\State \texttt{op := (T2(P,i,j,k),(j,k))} \Comment{T2 on qubits $(j,k)$ reduces  $F_{ij},F_{ik}$ to rank 1}
\State \texttt{applyOp(P,op), opList.append(op)}
\EndFor
\State \texttt{R := [j for j in [0..n-1] if Rank(F(P,i,j) = 1]}
\For{\texttt{j in R}}
\State \texttt{op := (T1(P,i,p,j),(i,j))} \Comment{T1  on qubits $(i,j)$ reduces $F_{ij}$ to zero}
\State \texttt{applyOp(P,op), opList.append(op)}
\EndFor
\EndFor
\State \texttt{return P, opList}
\end{algorithmic}

\subsection{Algorithms for Optimising Circuit Depth}
Our algorithms can be used to either optimise for entangling 2-qubit gate count or circuit depth.
We now describe two important helper functions to facilitate depth optimisation.
The first is a function for determining the depth of a circuit.
Circuits are stored as lists of ordered pairs \texttt{(opType,qList} where \texttt{opType} is a string describing the gate type and \texttt{qList} is a list of qubit indices to which the gate is applied:
\begin{algorithm}[H]
\caption{depth(opList)}\label{alg:depth-function}
\end{algorithm}
\vspace{-4ex}
\begin{algorithmic}
\State\textbf{Input:} 
\State \texttt{opList}: a list of gates
\State\textbf{Output:}  
\State \texttt{d}: depth of the circuit corresponding to \texttt{opList}.
\State\textbf{Method:} 
\State \texttt{layers := []}
\For{\texttt{(opType,qList) in opList}}
\If{\texttt{opType not in [`QPerm',`SWAP'] and len(qList) > 1}}
\State \texttt{L:=len(layers), i := L}
\While{\texttt{i > 0  and (qList $\cap$ layers[i-1]) != $\emptyset$}}
\State \texttt{i := i - 1}
\EndWhile
\If{\texttt{i = L}}:
\State \texttt{layers.append(qList)}
\Else
\State \texttt{layers[i] := layers[i] + qList}
\EndIf
\EndIf
\EndFor
\State \texttt{return len(layers)}
\end{algorithmic}

We now describe an algorithm which is used to find all circuits of depth one for a particular 2-qubit gate set.
The \texttt{QPart} function  is a recursive function \begin{added}that\end{added} returns all combinations of non-overlapping pairs of qubits from a set of qubit indices \texttt{S} which increase the depth of the circuit.
It takes as input the support \texttt{Supp} of the previous layer of gates.

\begin{algorithm}[H]
\caption{QPart(S, Supp)}\label{alg:depth-one}
\end{algorithm}
\vspace{-4ex}
\begin{algorithmic}
\State\textbf{Input:} 
\State \texttt{S}: a list of qubit indices
\State\textbf{Output:}  
\State \texttt{pList}: list of all possible sets of non-overlapping ordered pairs with elements from \texttt{S}
\State\textbf{Method:} 
\State \texttt{pList := [[]], LS = len(S)}
\For{\texttt{i in [0..LS-2]}}
\For{\texttt{j in [i+1..LS-1]}}
\State \texttt{p := (i,j)}
\If{\texttt{p $\cap$ Supp != $\emptyset$}}
\State \texttt{S1 = [S[k] for k in ([i+1..j-1,j+1..LS-1])]}
\State \texttt{pList := pList + [([p] + pList1) for pList1 in Qpart(S1)])}
\EndIf
\EndFor
\EndFor
\State \texttt{return pList}
\end{algorithmic}

\subsection{Greedy Synthesis Algorithm}
We next set out a greedy algorithm which generates a circuit using a scalar or vector heuristic (see \Cref{eq:H_heuristics,eq:h_vector}).
The algorithm can either be run for invertible matrices or symplectic matrices.
By setting \texttt{minDepth := TRUE}, the user can optimise for minimum depth rather than minimum 2-qubit gate-count.
The algorithm terminates if \texttt{maxWait} gates have been applied without reducing the optimisation heuristic, allowing us to track whether the algorithm has become trapped in a local minimum.
The function \texttt{applyOp(A,op)} applies a two-qubit gate  to the binary symplectic matrix \texttt{A}.

\begin{algorithm}[H]
\caption{Greedy Synthesis}\label{alg:greedy}
\end{algorithm}
% \end{figure}
\vspace{-4ex}
\begin{algorithmic}
\State\textbf{Input:} 
\State{\texttt{A}: a $2n\times 2n$ binary symplectic matrix}
% \State{\texttt{mode}: `GL' or `SP'}
\State{\texttt{gateOpts}: a function returning 2-qubit gate options which can be applied to a matrix}
\State{\texttt{Heuristic}: a heuristic function which is zero if the matrix is in the desired final form}
\State{\texttt{minDepth}: \texttt{TRUE} to find minimize depth or \texttt{FALSE} to minimize 2-qubit gate count.}
\State{\texttt{maxWait}: max iterations to try improving the heuristic before exiting;}
\State\textbf{Output:}  
\State{\texttt{P}: reduced matrix; AND}
\State{\texttt{opList}: series of 2-qubit operators yielding \texttt{A} when applied to \texttt{P}; OR}
\State{\texttt{FALSE}: if \texttt{maxWait} exceeded}
\State\textbf{Method:} 
\State \texttt{opList := [], P:= copy(A)}
\State \texttt{hLast := Heuristic(P), hMin := hLast}
\State \texttt{currWait := 0}
\While{\texttt{hLast != 0}}
\State \texttt{dhMin:=None}\Comment{find gate which minimises depth and heuristic}
\For{\texttt{op in GateOpts(A)}}
\State \texttt{B := applyOp(P,op)}
\State \texttt{hB := Heuristic(B)}
\State \texttt{dB := 0}
\If{\texttt{minDepth}}
\State \texttt{dB := depth(opList + [op])}
\If{\texttt{hB > hLast}}
\State \texttt{dB := dB + 10000}\Comment{penalise gates which do not improve heuristic}
\EndIf
\EndIf
\If{\texttt{dhMin is None or dhMin > (dB,hB)}}
\State \texttt{dhMin := (dB,hB), opMin := op}
\EndIf
\EndFor
\State \texttt{dLast, hLast := dhMin}\Comment{update \texttt{hLast, P} and \texttt{opList}}
\State \texttt{P := applyOp(P,opMin)}
\State \texttt{opList.append(opMin)}
\If{\texttt{hMin > hLast}}
\State \texttt{hMin := hLast, currWait := 0}  \Comment{better than \texttt{hMin} - reset \texttt{currWait}}
\Else
\State \texttt{currWait := currWait+1}\Comment{\texttt{hLast} worse than \texttt{hMin} - increment \texttt{currWait}}
\EndIf
\If{\texttt{currWait > maxWait}}\Comment{exit if \texttt{maxWait} gates since heuristic improved}
\State return False
\EndIf
\EndWhile
\State \texttt{return inverse(opList), P}
\end{algorithmic}

\subsection{A* Synthesis Algorithm}
Here we set out the A* synthesis algorithm of \Cref{sec:CNOT_astar,sec:clifford_astar}.
The A* algorithm takes one of the heuristics of \Cref{eq:H_heuristics} as input.
The algorithm can either be run for invertible matrices or symplectic matrices.
By setting \texttt{minDepth := TRUE}, the user can optimise for minimum depth rather than minimum 2-qubit gate-count.
The \texttt{GateOpts} function gives the possible gates which can be applied to the matrix at each step.
The size of the priority queue is limited to the input \texttt{maxQ} - we use the treap data structure and \texttt{removeMax} function to manage the queue length. 
We also maintain a tree database in \texttt{TreeDB}. 
Each row represents a matrix and stores only the parent ID plus the gate required to reach the corresponding matrix to reduce memory overhead. 
We use the function \texttt{TreeDB.retrieve(AId)} to retrieve   matrix representation and the gate list of  the row indexed by \texttt{AId}. 

\begin{algorithm}[H]

\caption{A* Synthesis}\label{alg:astar_synthesis}
\end{algorithm}
% \end{figure}
\vspace{-4ex}
\begin{algorithmic}
\State\textbf{Input:} 
\State \texttt{A}: a $2n\times 2n$ binary symplectic matrix
\State{\texttt{gateOpts}: a function returning 2-qubit gate options which can be applied to a matrix}
\State{\texttt{Heuristic}: a heuristic function which is zero if the matrix is in the desired final form}
\State{\texttt{minDepth}: \texttt{TRUE} to find minimize depth or \texttt{FALSE} to minimize 2-qubit gate count.}
\State \texttt{maxQ}: maximum size of the priority queue.
\State\textbf{Output:}  
\State \texttt{B}: the final reduced matrix; AND
\State{\texttt{opList}: series of 2-qubit operators yielding \texttt{A} when applied to \texttt{B}}
\State\textbf{Method:} 
\State{\texttt{B := copy(A),AId :=-1,BId:=0,op:= None,gB:=0,hB:=Heuristic(B)}}
\State \texttt{visited:=[B]}
\State \texttt{TreeDB := [(AId,op)]}\Comment{parent and gate.}
\State \texttt{PriQ := new treap()}
\State \texttt{PriQ.append((gB+hB,gB,BId))} \Comment{Insert item with priority \texttt{gB+hB} to priority queue}
\While{\texttt{len(todo) > 0}}
\State \texttt{ghA,gA,AId := PriQ.removeMin()}
\State\texttt{A, opListA:=TreeDB.retrieve(AId)}
\For{\texttt{op in GateOpts(A)}}
\State \texttt{B := applyOp(A,op)}
\If{\texttt{B not in visited}}
\State \texttt{hB := Heuristic(B), BId := BId + 1}
\State \texttt{TreeDB.append((AId,op))}
\If{\texttt{hB = 0}}
\State \texttt{return B, opListA + [op]}
\EndIf
\State \texttt{gB := depth(opListA + [op]) if minDepth else gA+1}
\State \texttt{PriQ.append((gB+hB,gB,BId)), visited.append(B)}
\EndIf
\EndFor
\While{\texttt{len(PriQ) > maxQ}}\Comment{remove excess items from \texttt{PriQ}}
\State \texttt{PriQ.removeMax()}
\EndWhile
\EndWhile
\State\texttt{return FALSE}
\end{algorithmic}

\subsection{Sensitivity Analysis for A*}
Here we present results on optimising the heuristic $h$ for the A* algorithm. We consider both changing the parameter $r$ and swapping between the $H_\text{sum}$ and $H_\text{prod}$ metrics of \Cref{eq:H_heuristics}.

\begin{figure}[H]
    \centering
    \includegraphics[width=0.7\linewidth]{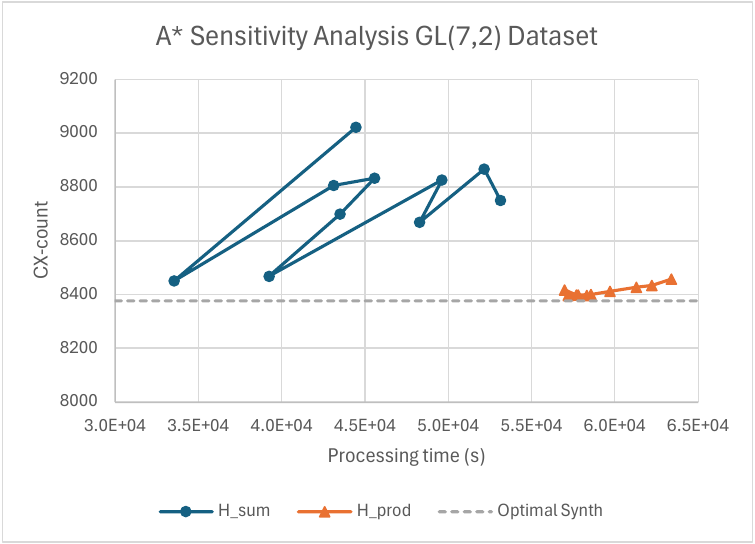}
    \caption{Sensitivity analysis for \CNOT A* synthesis algorithm. The blue line represents A* using the matrix weight heuristic $H_\text{sum}(A)$ for values of $r$ in the range 2.0-3.0. The orange line uses the log of col and row sums heuristic $H_\text{prod}(A)$ for $r$ in the range 1.5-2.4. We plot the total \CNOT-count for the data set of 970 elements of $\GL(7,2)$ versus the total time taken. We see that $H_\text{prod}(A)$ heuristic gives results closer to the known optimal \CNOT-count of 8376 for the data set.}
    \label{fig:CNOT_astar_sensitivity}
\end{figure}

For the \CNOT circuit case, a sensitivity analysis performed on the $\GL(7,2)$ dataset for the different cost function heuristics gives us the following results in Figure \ref{fig:CNOT_astar_sensitivity}, with data \href{https://github.com/m-webster/CliffordOpt/blob/main/paper_results/GL7-sensitivities.xlsx}{available online}.

\begin{figure}[H]
    \centering
    \includegraphics[width=0.7\linewidth]{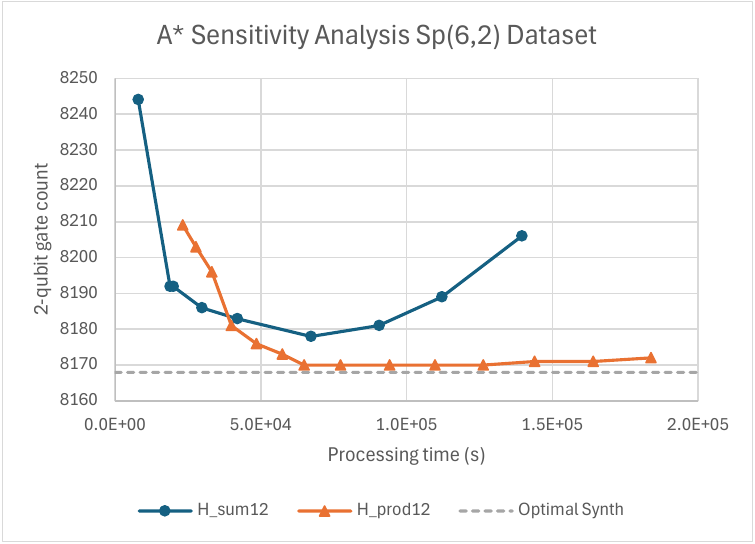}
    \caption{Sensitivity analysis for Clifford A* synthesis algorithm. The blue line represents A* using the matrix weight heuristic $rH_\text{sum}(A)$ for values of $r$ in the range 1.65-2.05. The orange line uses the log of col and row sums heuristic $rH_\text{prod}(A)$ for $r$ in the range 3.7-5.0. We plot the total two-qubit gate-count for the data set from \cite{Bravyi_Shaydulin_Hu_Maslov_2021} of 1003 elements of $\Sp(6,2)$ versus the total time taken. We see that $H_\text{prod}(A)$ heuristic gives results closer to the best-known  gate-count of 8168 for the data set.}
    \label{fig:Sp_astar_sensitivity}
\end{figure}

Similarly for the Clifford circuit synthesis, using the $6$ qubit Clifford circuit dataset of \cite{Bravyi_Shaydulin_Hu_Maslov_2021}, we obtain the below results in Figure \ref{fig:Sp_astar_sensitivity}, with data \href{https://github.com/m-webster/CliffordOpt/blob/main/paper_results/Sp6-sensitivities.xlsx}{available online}.

\subsection{Generating a Database of Matrix Equivalence Classes}
We now set out the algorithm for generating a database of equivalence classes of matrices in $\GL(n,2)$ or $\Sp(2n,2)$. 
As in the A* algorithm, we save only the tree structure of the database and retrieve the gate list and matrix using the \texttt{TreeDB.retrieve(AId)} function.
Gate options are calculated using the \texttt{GateOpts} function.
For optimal depth synthesis, \texttt{GateOpts} returns the set of all depth-one circuits calculated using the method of \Cref{alg:depth-one}.
Otherwise,  \texttt{GateOpts} returns the set of all two-qubit gate options - for \CNOT synthesis these are all possible \CNOT gates on $n$ qubits, whereas for Clifford synthesis it is the set of all two-qubit transvections (see \Cref{sec:transvections}).
The function \texttt{Canonize(B)} generates the graph corresponding to B and returns the certificate for the equivalence class.

\begin{algorithm}[H]
\caption{Optimal Database Generation}\label{alg:optimal_DB}
\end{algorithm}
% \end{figure}
\vspace{-4ex}
\begin{algorithmic}
\State\textbf{Input:} 
\State{\texttt{n}: number of qubits}
\State{\texttt{gateOpts}: function giving possible gates which can be applied to a matrix. For optimal depth mode, these are  depth-one circuits.}
\State{\texttt{Canonize}: function giving canonical graph isomorphism certificate for a matrix.}
\State\textbf{Output:}  
\State \texttt{TreeDB}: A database where each row corresponds to an equivalence class of matrices and gives  the parent node and the operation required to generate the matrix.
\State\textbf{Method:} 
\State \texttt{B:=I(2n)}
\State \texttt{BCert:=Canonize(B)}
\State \texttt{visited:=[BCert]}\Comment{list of visited equivalence classes}
\State \texttt{AId:=None, op:=None}\Comment{No parent or operation for identity matrix}
\State \texttt{TreeDB:=[(AId,op,0,0)]}\Comment{Tree database - store parent and operation in a list}
\State \texttt{BId := 0}\Comment{pointer to last row of TreeDB}
\State \texttt{SId := 0}\Comment{pointer to first row of current iteration}
\While{\texttt{BId >= SId}}
\State \texttt{SId := BId + 1}
\For{\texttt{AId in [SId..BId]}}\Comment{iterate through newly added rows of \texttt{TreeDB}}
\State \texttt{A,opListA:= TreeDB.retrieve(AId)}
\For{\texttt{inv in [0,1]}}
\For{\texttt{transp in [0,1]}} \Comment{apply gates to $A,A^{-1},A^T,A^{-T}$}
\For{\texttt{op in GateOpts(A)}}\Comment{apply all possible gates}
\State \texttt{B:=applyOp(A,op)}
\State \texttt{BInv:=Inverse(B)}
\State \texttt{BCert := min([Canonize(Bi) for Bi in [B, B.T, BInv, BInv.T]])}
\If{\texttt{BCert not in visited}}
\State \texttt{TreeDB.append((AId, op, transp, inv))}\Comment{add new row to \texttt{TreeDB }}
\State \texttt{visited.append(BCert)}\Comment{update \texttt{visited}}
\State \texttt{BId := BId + 1}\Comment{increment \texttt{BId}}
\EndIf
\EndFor
\EndFor
\State \texttt{A:=Transpose(A)}
\EndFor 
\State \texttt{A:=Inverse(A)}
\EndFor
\EndWhile
\State \texttt{return TreeDB}
\end{algorithmic}
\begin{added}
\section{Equivalence of Clifford Operators up to Single-Qubit Cliffords and Permutations}\label{sec:graph_iso_proof}
In this section we explain why the method in \Cref{fig:sym_graph} correctly identifies equivalence classes of symplectic matrices up to qubit permutations and single qubit Cliffords acting on the inputs and output qubits.
For this purpose, we use a three-block symplectic form of a Clifford operator $A$ defined as follows:
\begin{align}
    \Sp(A) &:= \begin{bmatrix}A_{XX}&A_{XZ}\\A_{ZX}&A_{ZZ}\end{bmatrix},\\\text{ESp}(A) &:=  \begin{bmatrix}A_{XX}&A_{XZ} & A_{XX}\oplus A_{XZ}\\A_{ZX}&A_{ZZ}& A_{ZX} \oplus A_{ZZ} \\ A_{XX} \oplus A_{ZX} &A_{XZ} \oplus A_{ZZ} & A_{XX}\oplus A_{XZ} \oplus A_{ZX} \oplus A_{ZZ}\end{bmatrix}.
\end{align}
First, we show that multiplication of a symplectic matrix by a qubit permutation corresponds to either row or column swaps within the same block of the 3-block symplectic matrix.
\begin{lemma}\label{lemma:transp}
    When acting the right of a Clifford operator, qubit transpositions swap the same two columns of the 3-block symplectic form of the Clifford within each block. When acting on the left, they swap the same rows within each block.
\end{lemma}
\begin{proof}
Consider a transposition $T$  acting on the identity Clifford operator on two qubits. This operator sends $X_0 \leftrightarrow X_1$ and $Z_0 \leftrightarrow Z_1$ and so has matrix representation:
\begin{align}
    \Sp(T) = \begin{bmatrix}0&1&0&0\\1&0&0&0\\0&0&0&1\\0&0&1&0\end{bmatrix} = \begin{bmatrix}T&0\\0&T\end{bmatrix},
\end{align}
where $T:=\begin{bmatrix}0&1\\1&0\end{bmatrix}$. When acting on the right of a symplectic matrix $\Sp(A)$,  $\Sp(T)$ swaps the columns within the first block of $\Sp(A)$ and the columns within the second block of $\Sp(A)$. When acting on the left, it swaps the rows of the first block of $\Sp(A)$ and the second block of $\Sp(A)$.

The 3-block symplectic matrix $\text{ESp}(T)$ swaps columns corresponding to the same qubits within the first, second and third block of $\text{ESp}(A)$ because it is of form:
\begin{align}
    \text{ESp}(T) = \begin{bmatrix}T&0&0\\0&T&0\\0&0&T\end{bmatrix}.
\end{align}
Because any permutation can be written as a product of transpositions, the same result follows for any qubit permutation.
\end{proof}

We now show that single qubit Cliffords acting on the right/left permute columns/rows corresponding to the same qubit between blocks of the 3-block symplectic matrix:
\begin{lemma}\label{lemma:SQC}
    When acting on the right of a Clifford operator, single-qubit Cliffords swap columns of the 3-block symplectic form of the Clifford corresponding to the same output qubit. When acting on the left, they swap rows corresponding to the same input qubit.
\end{lemma}
\begin{proof}
The $S$ operator corresponds to the symplectic matrix $\Sp(S) := \begin{bmatrix}1&1\\0&1\end{bmatrix}$ .
Assume that $S$ acts on the $i$th qubit and let $\mathbf{c}_{XX}$ be the $i$th column of the $XX$ component of the 3-block matrix. When acting on the right, $S$ has the following action on the $i$th columns of the blocks of the 3-block symplectic form:
\begin{align}
    \Sp(A)|_{\text{col }i}\Sp(S)  &= \begin{bmatrix}\mathbf{c}_{XX}&\mathbf{c}_{XZ} \\\mathbf{c}_{ZX}&\mathbf{c}_{ZZ} \end{bmatrix}\begin{bmatrix}1&1\\0&1\end{bmatrix}\\&=\begin{bmatrix} \mathbf{c}_{XX} &\mathbf{c}_{XX} \oplus \mathbf{c}_{XZ}   \\ \mathbf{c}_{ZX}&\mathbf{c}_{ZX} \oplus \mathbf{c}_{ZZ} \end{bmatrix}
\end{align}
Writing the result in 3-block symplectic form, we see that $S$ acting on the right swaps columns corresponding to the same qubit in the second and third block:
\begin{align}
    (\text{Sp}_3(A)|_{\text{col }i}) \text{Sp}_3(S)  &= \begin{bmatrix} \mathbf{c}_{XX} &\mathbf{c}_{XX} \oplus \mathbf{c}_{XZ} & \mathbf{c}_{XZ} \\ \mathbf{c}_{ZX}&\mathbf{c}_{ZX} \oplus \mathbf{c}_{ZZ} &\mathbf{c}_{ZZ}\\\mathbf{c}_{XX}\oplus \mathbf{c}_{ZX}&\mathbf{c}_{XX} \oplus \mathbf{c}_{XZ}\oplus \mathbf{c}_{ZX} \oplus \mathbf{c}_{ZZ} &\mathbf{c}_{XZ} \oplus \mathbf{c}_{ZZ}\end{bmatrix}
\end{align}
We now show that $S$ acting on the left corresponds to swapping rows. Let $\mathbf{r}_{XX}$ be the $i$th row of the $XX$ block of $A$:
\begin{align}\Sp(S)\Sp(A)|_{\text{row }i}   &= \begin{bmatrix}1&1\\0&1\end{bmatrix}\begin{bmatrix}\mathbf{r}_{XX}&\mathbf{r}_{XZ} \\\mathbf{r}_{ZX}&\mathbf{r}_{ZZ} \end{bmatrix}\\&=\begin{bmatrix}\mathbf{r}_{XX} \oplus \mathbf{r}_{ZX} & \mathbf{r}_{XZ} \oplus \mathbf{r}_{ZZ} \\ \mathbf{r}_{ZX}&\mathbf{r}_{ZZ} \end{bmatrix}.\end{align}
This action corresponds to swapping rows from the first and third blocks of the 3-block symplectic form.

Similarly, the Hadamard operator has the symplectic form $\Sp(H) := \begin{bmatrix}0&1\\1&0\end{bmatrix}$. A similar exercise demonstrates that Hadamard acting on the right swaps columns from the first and second blocks of the 3-block symplectic form and acting on the left swaps rows from the first and second blocks.
Because the single-qubit Clifford group is generated by $S$ and $H$, the result follows. \end{proof}

We now link the 3-block symplectic matrix representation back to the graph isomorphism problem and prove the main result of this section.

\begin{proposition}
    Two Clifford operators $A$ and $B$ acting on n qubits are equivalent up to qubit permutations and single qubit Cliffords acting on the left and right if and only if the graphs set out in \Cref{fig:sym_graph} are isomorphic.
\end{proposition}
\begin{proof}
Because $G_A$ has nodes of two colours (one for the columns of $\text{Sp}_3(A)$ and one for the rows) $G_A$ is isomorphic to $G_B$ if and only if there exists a row permutation $P$ and column permutation $Q$ such that $P\text{Sp}_3(A)Q = \text{Sp}_3(B)$ and the edges between the columns of $G_A$ (`column edges’) are sent to edges between the columns of $G_B$ and similarly for the ‘row edges’.

A column transposition which swaps the same columns within each block preserves the connectivity of the column edges, as does a column transposition which swaps columns corresponding to the same output qubit in different blocks. It is also easy to see that any column permutation which preserves the column edge connectivity must be a product of such transpositions. Due to \Cref{lemma:SQC,lemma:transp}, these transpositions correspond to either qubit permutations or single-qubit Clifford operators acting on the right, and the product of such transpositions forms the group of single-qubit Cliffords and permutations on $n$ qubits.

Similarly, the ‘row edges’ of $G_A$ are preserved by the group of single-qubit Cliffords and permutations acting on the left.

If in addition $P\text{Sp}_3(A)Q = \text{Sp}_3(B)$ then $B$ has the same symplectic form as $A$ up to single qubit Cliffords and permutations acting on the right and left and the result follows.
\end{proof}
\end{added}
\section{Equivalence Classes of Matrices by Gate-Count and Circuit Depth}
Here we include supplementary data on the equivalence classes of invertible and symplectic matrices grouped by minimum gate-count and circuit depth.

In \Cref{tab:GL2_CNOT_count,tab:GL2_depth} we list the equivalence classes of $\GL(n,2)$ by \CNOT count and circuit depth as generated by the optimal \CNOT synthesis algorithm of \Cref{sec:CNOT_optimal}.
\begin{table}[h]
    \centering
    \begin{tabular}{|r|r|r|r|r|r|r|}
    \hline
\CNOT Count    & n=2 & 3 & 4 & 5 & 6 & 7 \\
   \hline
 0 & 1 & 1 & 1 & 1 & 1 & 1 \\
 1 & 1 & 1 & 1 & 1 & 1 & 1 \\
 2 &  & 2 & 3 & 3 & 3 & 3 \\
 3 &  & 1 & 8 & 10 & 11 & 11 \\
 4 &  &  & 10 & 40 & 52 & 54 \\
 5 &  &  & 3 & 87 & 257 & 308 \\
 6 &  &  & 1 & 106 & 1,123 & 2,228 \\
 7 &  &  &  & 32 & 3,235 & 14,733 \\
 8 &  &  &  & 4 & 4,698 & 78,679 \\
 9 &  &  &  &  & 2,167 & 291,836 \\
 10 &  &  &  &  & 209 & 625,945 \\
 11 &  &  &  &  & 3 & 571,160 \\
 12 &  &  &  &  & 1 & 132,375 \\
 13 &  &  &  &  &  & 2,172 \\
 14 &  &  &  &  &  &2\\
 \hline
 Total & 2 & 5 & 27 & 284 & 11,761 & 1,719,491\\
  \hline
    \end{tabular}
    \caption{Classification of $\GL(n,2)$ matrix equivalence classes by minimum  \CNOT-count.}
    \label{tab:GL2_CNOT_count}
\end{table}

\begin{table}[h]
    \centering
    \begin{tabular}{|r|r|r|r|r|r|r|}
    \hline
Depth   & n=2 & 3 & 4 & 5 & 6 & 7 \\
   \hline
0&	1&	1&	1&	1&	1&	1\\
1&	1&	1&	2&	2&	3&	3\\
2&	&	2&	9&	17&	47&	91\\
3&	&	1&	14&	139&	1,805&	15,861\\
4&	&	&	1&	124&	9,687&866,885	\\
5&	&	&	&	1&	218&	\\
6&	&	&	&	&	&	\\
7&	&	&	&	&	&	\\
 \hline
Total&	2&	5&	27&	284&	11,761&	1,719,491\\
  \hline
    \end{tabular}
    \caption{Classification of $\GL(n,2)$ matrix equivalence classes by minimum circuit depth.}
    \label{tab:GL2_depth}
\end{table}

In \Cref{tab:Sp2_gate_count,tab:Sp2_depth} we list the equivalence classes of $Sp(2n,2)$ by 2-qubit gate count and circuit depth as generated by the optimal Clifford synthesis algorithm of \Cref{sec:clifford_optimal}.
We have calculated these up to $n=5$, and have partial results for $n=6$.
\begin{table}[h]
    \centering
    \begin{tabular}{|r|r|r|r|r|r|}
    \hline
 2-Qubit Gate Count & n=2 & 3 & 4 & 5 & 6 \\
\hline
 0 & 1 & 1 & 1 & 1 & 1 \\
 1 & 1 & 1 & 1 & 1 & 1 \\
 2 &  & 2 & 3 & 3 & 3 \\
 3 &  & 3 & 11 & 13 & 14 \\
 4 &  & 1 & 37 & 78 & 89 \\
 5 &  &  & 47 & 530 & 823 \\
 6 &  &  & 9 & 3,325 & 10,502 \\
 7 &  &  &  & 10,317 & 142,847 \\
 8 &  &  &  & 6,125 & 1,767,694 \\
 9 &  &  &  & 28 & 14,875,711 \\
 10 &  &  &  &  &  55,870,682 \\
 11 &  &  &  &  &  \\
 12 &  &  &  &  &  \\
 \hline
 Total & 2 & 8 & 109 & 20,421 & \\
  \hline
    \end{tabular}
    \caption{Classification of $\Sp(2n,2)$ matrix equivalence classes by minimum number of 2-qubit gate-count.}
    \label{tab:Sp2_gate_count}
\end{table}

\begin{table}[h]
    \centering
    \begin{tabular}{|r|r|r|r|r|r|}
    \hline
Depth & n=2 & 3 & 4 & 5 & 6 \\
\hline
0&	1&	1&	1&	1&	1\\
1&	1&	1&	2&	2&	3\\
2&	&	2&	11&	19&	54\\
3&	&	3&	84&	958&	2,071\\
4&	&	1&	11&	16,385&	62,731\\
5&	&	&	&	3,056&	\\
6&	&	&	&	&	\\
7&	&	&	&	&	\\
\hline
Total&	2&	8&	109&	20,421&	\\
  \hline
    \end{tabular}
    \caption{Classification of $\Sp(2n,2)$ matrix equivalence classes by circuit depth.}
    \label{tab:Sp2_depth}
\end{table}

\end{document}